\documentclass{article}

\usepackage{PRIMEarxiv}
\usepackage{comment}
\usepackage[utf8]{inputenc} %
\usepackage[T1]{fontenc}    %
\usepackage{hyperref}       %
\usepackage{url}            %
\usepackage{booktabs}       %
\usepackage{amsfonts}       %
\usepackage{nicefrac}       %
\usepackage{microtype}      %
\usepackage{lipsum}
\usepackage{fancyhdr}       %
\usepackage{graphicx}       %
\usepackage{subcaption}
\usepackage{amsmath} 
\usepackage{tabularx}
\usepackage{caption}       %
\graphicspath{{figures/}}     %
\usepackage[table,xcdraw]{xcolor}
\usepackage{array}
\usepackage{colortbl}
\usepackage{placeins}
\usepackage{soul,color}
\usepackage{multirow}
\usepackage{threeparttable}

\title{In context learning Foundation models for Materials Property Prediction with Small datasets
\thanks{\textit{\underline{Citation}}: 
\textbf{Li, et al.. ICL Foundation model for material property prediction. 16 Pages.... DOI:000000/11111.}} 
}

\author{%
  \textbf{Qinyang Li}$^{1}$ \footnotemark[3] \and
  \textbf{Rongzhi Dong}$^{1}$ \footnotemark[3] \and
  \textbf{Nicholas Miklaucic}$^{1}$ \and
  \textbf{Jeffrey Hu}$^{2}$ \and
  \textbf{Sadman Sadeed Omee}$^{1}$ \and
  \textbf{Lai Wei}$^{1}$ \and
  \textbf{Sourin Dey}$^{1}$ \and
  \textbf{Ming Hu}$^{3}$ \and
  \textbf{Jianjun Hu}$^{1}$\thanks{Corresponding author: jianjunh@cse.sc.edu} \\
  \\
  $^{1}$Department of Computer Science and Engineering, University of South Carolina, Columbia, SC, USA \\
  $^{2}$Department of Materials Science \& Engineering, University of Illinois Urbana-Champaign,IL, USA \\
  $^{3}$Department of Mechanical Engineering, University of South Carolina, Columbia, SC, USA \\
}

\begin{document}

\renewcommand{\thefootnote}{\fnsymbol{footnote}}
\footnotetext[3]{Equal contribution}
\maketitle

\begin{abstract}

Foundation models (FMs) have recently shown remarkable in-context learning (ICL) capabilities across diverse scientific domains. In this work, we introduce a unified in-context learning foundation model (ICL-FM) framework for materials property prediction that integrates both composition-based and structure-aware representations. The proposed approach couples the pretrained TabPFN transformer with graph neural network (GNN)-derived embeddings and our novel MagpieEX descriptors. MagpieEX augments traditional features with cation-anion interaction data to explicitly measure bond ionicity and charge-transfer asymmetry, capturing interatomic bonding characteristics that influence vibrational and thermal transport properties.
Comprehensive experiments on the MatBench benchmark suite and a standalone lattice thermal conductivity (LTC) dataset demonstrate that ICL-FM achieves competitive or superior performance to state-of-the-art (SOTA) models with significantly reduced training costs. Remarkably, the training-free ICL-FM outperformed sophisticated SOTA GNN models in five out of six representative composition-based tasks, including a significant 9.93\% improvement in phonon frequency prediction. On the LTC dataset, the FM effectively models complex phenomena such as phonon-phonon scattering and atomic mass contrast. t-SNE analysis reveals that the FM acts as a physics-aware feature refiner, transforming raw, disjoint feature clusters into continuous manifolds with gradual property transitions. This restructured latent space enhances interpolative prediction accuracy while aligning learned representations with underlying physical laws. This study establishes ICL-FM as a generalizable, data-efficient paradigm for materials informatics.

\end{abstract}

\section*{Introduction}

The discovery and design of novel materials are fundamental for developing energy storage, semiconductor technology, catalysis, and numerous other industrial applications, %
critical to addressing global challenges and fostering transformative progress in modern society. 
A central bottleneck in the material discovery process is accurate and fast prediction of fundamental materials properties such as ion conductivity, thermal conductivity, formation energy, band gap, dielectric constants, elastic moduli, etc. Such material property prediction models have been widely used in screening new functional materials from known materials repositories \cite{ojih2024graph}, inverse materials design \cite{zeni2025generative}, and guiding experimental explorations \cite{jin2023atomic}.
While quantum-mechanical simulations such as density functional theory (DFT) 
may achieve accurate predictions for some materials properties, they are usually computationally expensive and cannot be scaled to the millions of candidate compounds required for high-throughput virtual screening. This has motivated the rapid development of machine learning (ML) methods for materials property prediction as a complementary approach to accelerate materials discovery. 

In the past decade, machine-learning-based materials property prediction models
have achieved significant progress \cite{dunn2020benchmarking} and are increasingly integrated into materials discovery pipelines \cite{wang2022benchmarking,rohr2020benchmarking}. These include composition-based property prediction algorithms based on the widely used Magpie (2016) features  \cite{ward2016general}, recent 
attention-based neural network Roost (2020) \cite{goodall2020predicting} and CrabNet (2021) \cite{wang2021compositionally}, neural network based ModNet (2021) \cite{de2021materials}, and large-language-model-based Darwin (2023) \cite{xie2023large}, which learn to quickly predict material properties directly from the composition, making them suitable for high-throughput screening candidate materials without structural information. 

Compared to composition-based predictors, structure-based material property prediction models have demonstrated higher performance and stronger generalization across diverse property datasets, especially those graph neural networks (GNN) based models \cite{reiser2022graph} from early CGCNN \cite{xie2018crystal} to more recent ALIGNN \cite{choudhary2021atomistic}, DeeperGATGNN\cite{omee2022scalable}, M3GNet\cite{chen2022universal}, coGN, coNGN \cite{ruff2023connectivity}, and CHGNet \cite{deng_2023_chgnet}. These models incorporate atomic connectivity and geometric features to extract better structural and physicochemical features from the structures for accurate property prediction. A series of architectural innovations in neural networks have been introduced to model multi-body interactions and bond angles \cite{chen2022universal,choudhary2021atomistic,hsu2022efficient,gasteiger2021gemnet}. To deal with periodicity \cite{yan2022periodic} and transformations such as translation, rotation, and reflection using equivariant neural networks \cite{gasteiger2021gemnet,batzner20223,liao2022equiformer,yan2022periodic}, Transformers \cite{crabnet,madani2025accelerating}, graph transformers \cite{gasteiger2021gemnet,yan2022periodic,tao2025cgformer}, transfer learning \cite{madani2025accelerating,gupta2021cross}, active learning \cite{jose2024informative}, self-supervised learning and the pretraining+finetuning approaches \cite{fu2024physics}, and multi-modal neural networks \cite{moro2025multimodal,chen2025multi}, have been proposed to improve material property prediction.

Despite substantial research progress, the challenge of accurately predicting material properties from small datasets remains unsolved. This is evident from the stagnation of prediction performance on the Matbench \cite{dunn2020benchmarking} leaderboards since 2023, where most mean absolute error (MAE) improvements of state-of-the-art (SOTA) algorithms over ALIGNN (2021) \cite{choudhary2021atomistic} are below 10\%. In a separate study, the PMCGNN (2025) \cite{feng2024improving} model, which combines global and local modules, reported an MAE of 0.038 GPa for bulk moduli prediction (a 5.00\% reduction compared to PotNet \cite{lin2023efficient}, 2023) and 0.063 GPa for shear moduli prediction (a 3.08\% reduction compared to PotNet, 2023). In another study of 2025 \cite{madani2025accelerating}, their sophisticated CrysCo algorithm achieves similar performance (with <3.5\% improvement) for six out of eight datasets compared to ALIGNN (2021). Their transfer learning model CrysCoT also only achieved marginal improvement (<3\%) compared to CrysCo as it is easy to overfit the target datasets during the finetuning stage. Similar marginal improvements can also be found in the multi-modal model \cite{moro2025multimodal}. 

Most materials property prediction tasks with high-demand industrial applications are inherently defined over small datasets such as ionic conductivity \cite{datta2022conductivity}, thermal conductivity \cite{ojih2023screening}, elastic moduli \cite{dunn2020benchmarking}, piezoelectric tensors \cite{dong2025accurate}, and dielectric properties \cite{dunn2020benchmarking}, usually containing a few hundreds to a few hundreds of samples. The complexity of the structure-property relationship coupled with such small datasets and GNNs' high expressive power naturally leads to overfitting: the model learns the training examples including its noises instead of learning general rules. This weakness is further compounded by inherent material dataset redundancy \cite{li2024md,hossain2025surprisingly,li2023exploiting} and the challenge for out-of-distribution predictions needed for discovering novel outlier/exceptional materials \cite{omee2024structure,li2025probing}

Recently, the broader field of machine learning has been transformed 
by the emergence of foundation models (FMs) \cite{fei2022towards,chen2024towards,hao2024large,mishra2024foundational}. 
Unlike task-specific architectures, FMs are trained on massive and heterogeneous datasets using scalable architectures and objectives, 
producing general-purpose representations that can be transferred across diverse downstream tasks. This paradigm has reshaped diverse fields such as natural language processing \cite{min2023recent}, computer vision \cite{awais2025foundation}, medical imaging\cite{willemink2022toward}, and chemistry \cite{king2024transfer}. Foundation models can be used for addressing the persistent ML challenges such as data scarcity and poor generalization \cite{choi2025perspective}. In the field of materials science \cite{pyzer2025foundation}, large-language model (LLM) based foundation models have been applied to diverse materials discovery tasks such as data extraction, property prediction, and synthesis planning \cite{pyzer2025foundation}. However, it remains a challenge how to link such LLM models to highly specialized GNN models for materials property prediction. FMs trained with supervised pretraining over large-scale structure-formation energy data (1.58M) have recently become popular in machine learning potentials models such as MACE, SevenNet, eSEN as shown in the matbench-discovery leaderboard \cite{riebesell2023matbench}. Extract universal features from such models for generic property prediction is nontrivial as effective transfer depends on appropriate alignment between pretraining and target tasks. So far, both self-supervised learning based \cite{magar2022crystal} and multi-modality FMs \cite{moro2025multimodal} have yet to demonstrate significant performance improvement for small-dataset material property prediction.

These findings establish in context learning FMs \cite{hollmann2025accurate,qu2025tabicl} as a versatile and generalizable paradigm for scientific property prediction, highlighting both their opportunities and current limitations, and laying the groundwork for FM-enabled, plug-and-play pipelines to accelerate materials discovery.  

In summary, this work introduces a unified in-context learning foundation model (ICL-FM) framework for materials property prediction that bridges composition-based and structure-aware representations. By coupling the pretrained TabPFN transformer with both handcrafted Magpie descriptors and graph-based embeddings from ALIGNN and CGCNN, we demonstrate that foundation models can achieve state-of-the-art accuracy while providing interpretable insights into material-property relationships. Comprehensive evaluations across the MatBench benchmark suite and a standalone thermal conductivity dataset confirm the model’s robustness, generalization, and physical consistency. This study establishes ICL-FM as a versatile and data-efficient paradigm for advancing foundation-model applications in materials informatics.

\section*{Results}
\label{sec:results}
\subsection*{In-Context Learning Foundation Model based property predictors(ICL-FM)}
In-context learning (ICL) \cite{brown2020language,dong-etal-2024-survey} refers to a behavior of large transformers networks where they learn to perform a new task during inference purely from examples provided in the input, without any change to the model’s weights. 
ICL can be seen as parameterized models predict from analogy, which condition its specific predictive behavior on its inference-time prompts including zero-shot learning (with only task description), one-shot learning (with task description plus exactly one example), and few-shot learning (with a small set of examples). These prompts are used to help the model lock into the desired format or logic, essentially "locating" and activating specific knowledge it already possessed for the requested task. Another way to understand ICL is that the prompt creates a "task vector" in the model's activation space that shifts the output toward the desired behavior. A recent study shows that the emergent ICL capability is derived from the pretraining with diverse tasks \cite{lu2025asymptotic}, for which the task generalization capability of ICL can be linked to meta-learning as emergent or implicit metalearning \cite{genewein2025understanding}. 

The emergent general-purpose in-context learning can apply to many tasks—classification, translation, reasoning, function induction, even simple algorithm learning, enabling these models to have zero-shot and few-shot capabilities. Recently, a transformer-based foundation model designed for tabular data, TabPFN (Tabular Prior-Data Fitted Network), was proposed \cite{hollmann2025accurate}, which pre-trains a transformer network on millions of synthetically generated tabular datasets, achieving strong in-context learning performance on small datasets for both regression and classification tasks. Here we demonstrated that this model can surpass current sophisticated SOTA models for composition based material property prediction.

\begin{figure}[htbp]
    \centering
    \includegraphics[width=0.85\textwidth]{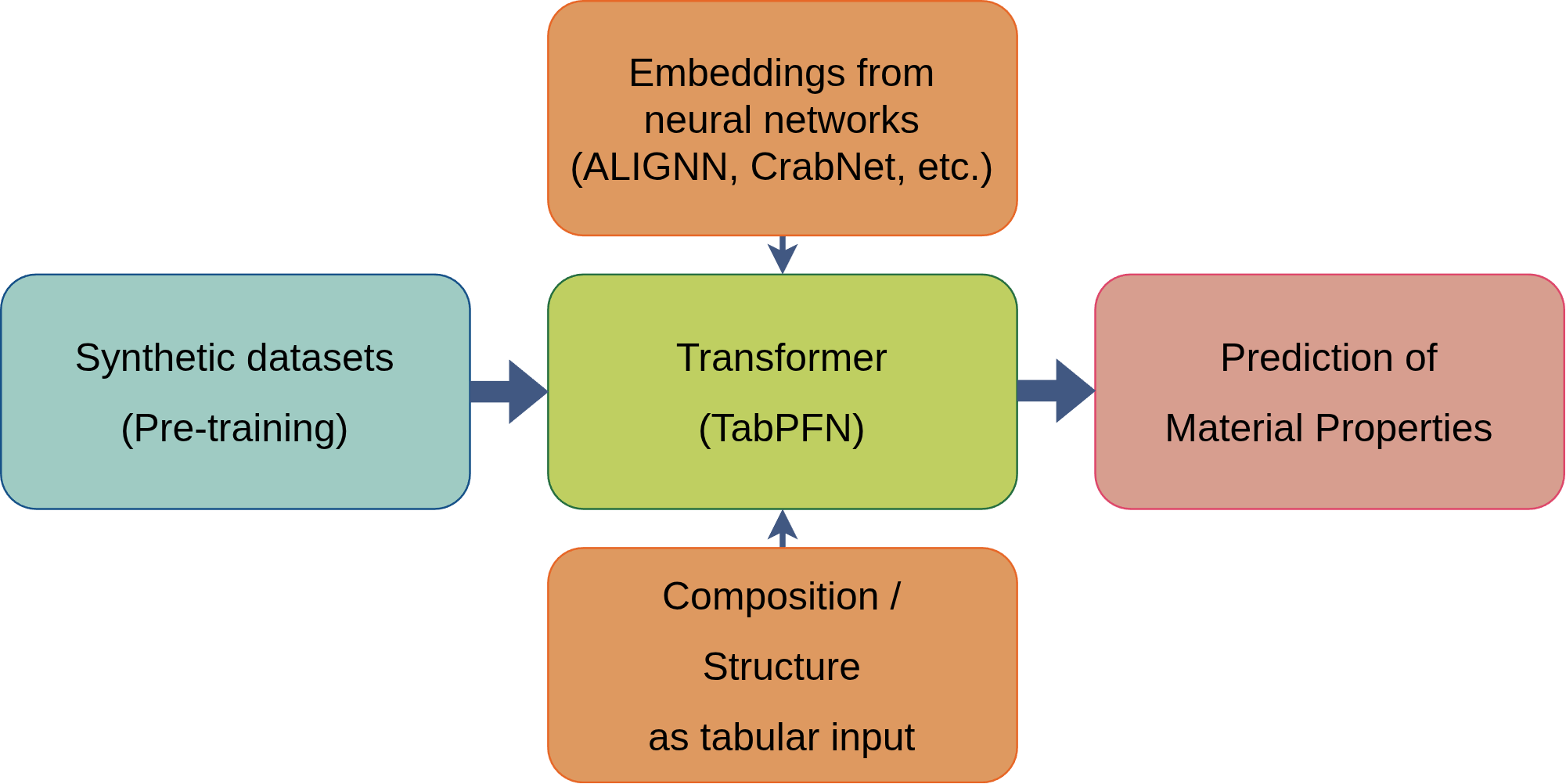}
    \caption{TabPFN based ICL-FM framework for materials property prediction: (left) The transformer based TabPFN model is pretrained using millions of synthetic regression datasets; (top/bottom): material representations can be composition and structural descriptors or embeddings extracted from GNN models; (right): material feature-property pairs are fed to the FM model for training-free in-context property prediction. }
    \label{fig:struct_bars}
\end{figure}

Figure~\ref{fig:struct_bars} illustrates the overall workflow of the proposed In-Context Learning Foundation Model (ICL-FM) for material property prediction. The backbone of the framework is the Transformer-based TabPFN model, which is pre-trained on millions of synthetically generated tabular datasets to acquire strong in-context learning capabilities. For materials applications, the model takes two types of representations as input: traditional composition features such as Magpie descriptors, or learned embeddings produced by neural network–based materials models (e.g., ALIGNN, CrabNet). These features are formatted as tabular inputs and processed by TabPFN as a small in-context regression task. Leveraging its pre-trained prior, the model can rapidly generalize to new materials without any weight updates, ultimately providing accurate predictions of target material properties. This framework highlights the potential of tabular foundation models for data-efficient materials informatics.

\paragraph*{TabPFN: A Foundation Model for Tabular Data}
TabPFN \cite{hollmann2025accurate} as first proposed in 2022 is a transformer model trained offline on millions of synthetically generated datasets using structural causal models. TabPFN learns a general-purpose in-context learning capability during the pretraining stage. At inference time, it processes both training and test samples in a single forward pass, requiring no additional training or hyperparameter tuning. Empirically, TabPFN has demonstrated competitive or superior performance compared to state-of-the-art machine learning algorithms, including boosted trees and AutoML systems on small- to medium-sized datasets (up to ~1,000 samples with v1; ~10,000 in later benchmarks), while offering orders-of-magnitude speedups. Mechanistically, TabPFN approximates Bayesian inference with a preferred prior over causal synthetics, allowing it to generalize effectively across new datasets with minimal overhead.

TabPFN’s small-data learning ability and its tabular data processing make it particularly appealing for materials informatics problems, where small datasets are common. It can naturally handle composition-based features or higher-level representations transformed into tabular form. In this study, we demonstrate that integrating TabPFN as a prediction head on top of structural embeddings extracted from graph-based models requires only minimal modification: the pseudo code in Supplementary Note Algorithm S1 shows that the entire ICL-FM framework operates in a plug-and-play manner with just a few lines of implementation. Despite this simplicity and the absence of model-specific finetuning, the approach achieves strong predictive performance, highlighting the practical usability of TabPFN for rapid materials property modeling.

\paragraph*{Magpie Features and Neural Network Embedding Representations}

To evaluate the performance of our proposed In-Context Learning Foundation Models (ICL-FM) for materials property prediction, we considered two complementary types of feature representations:
(1) composition-based handcrafted descriptors and
(2) latent embeddings extracted from neural networks trained on structural data.

The Magpie feature set~\cite{ward2016general} encodes fundamental elemental statistics, such as mean electronegativity, atomic radius, and valence electron count, that serve as a physically interpretable baseline for composition-based learning.
To extend Magpie’s chemical expressiveness, we introduce an enhanced variant, MagpieEX, which augments the original descriptors with cation–anion interaction features derived from Pymatgen’s oxidation-state analysis.
For each compound, we calculate average cation and anion attributes (electronegativity, atomic radius, valence electron configuration, ionization energy, electron affinity, and polarizability) and include their differences as explicit measures of bond ionicity and charge-transfer asymmetry.
This extension enables the model to capture interatomic bonding characteristics that influence properties such as lattice vibrations, dielectric response, and thermal transport.

We also tested the ICL-FM approach for the second materials representation, which leverages structural embeddings extracted from graph neural networks such as ALIGNN~\cite{choudhary2021atomistic} and CGCNN~\cite{xie2018crystal}. 
We perform \emph{feature embedding extraction} by capturing the activations from the penultimate (pre–fully-connected) layer of these pre-trained models. This process, also known as \emph{latent representation transfer}~\cite{pan2009survey,erhan2010understanding}, retains high-level structural correlations while abstracting away from the task-specific output mapping. 
These embeddings are then treated as tabular inputs to the TabPFN model, effectively allowing it to act as a plug-in predictor on top of learned structural representations. This hybrid approach combines structural expressiveness and fast, training-free inference, enabling a unified framework for both composition- and structure-based property prediction.

\subsection*{Datasets}
To systematically evaluate the ICL approach for materials property prediction, we conduct experiments on both standard benchmarks and a custom material dataset. We primarily use the \texttt{Matbench} suite~\cite{dunn2020benchmarking}, a standardized benchmark designed for fair assessment of machine learning models for materials property prediction Additionally, we include a curated thermal conductivity dataset to test the model’s generalization on thermophysical properties.

\paragraph*{Matbench Benchmark Dataset}
The \texttt{Matbench} suite is part of the Materials Project ecosystem and contains 13 regression and classification tasks derived from experimentally measured or density functional theory (DFT)-calculated properties. Each task provides fixed train/test splits to ensure reproducibility.  
In this study, we focus on six representative regression tasks covering diverse physical properties:
\begin{itemize}
    \item \textbf{Matbench\_jdft2d}: Exfoliation energy (eV/atom) prediction for 2D materials based on JDFT calculations (636 samples).
    \item \textbf{Matbench\_phonons}: Frequency of the highest frequency optical phonon mode peak (1/cm) prediction for crystalline materials (1,265 samples).
    \item \textbf{Matbench\_dielectric}: Logarithm of the refractive index prediction for inorganic compounds (4,764 samples).
    \item \textbf{Matbench\_log\_gvrh}: Logarithm of the shear modulus (GPa) prediction using Voigt–Reuss–Hill averaging (10,987 samples).
    \item \textbf{Matbench\_log\_kvrh}: Logarithm of the bulk modulus (GPa) prediction using Voigt–Reuss–Hill averaging (10,987 samples).
    \item \textbf{Matbench\_perovskites}: Formation energy (eV/atom) prediction for perovskite structures (18,928 samples).
\end{itemize}
These tasks cover a wide range of physical properties and dataset  sizes, allowing for robust comparisons between composition-based and structure-informed predictors.

\paragraph*{Thermal Conductivity Dataset}
To complement the benchmark tasks, we compiled an additional dataset for predicting lattice thermal conductivity. 
This dataset contains 3,149 unique compounds with room-temperature lattice thermal conductivity (LTC) values with both composition and structures. 
Since thermal conductivity sensitively depends on anharmonic lattice dynamics, this dataset provides an additional challenge for evaluating generalization beyond purely electronic or elastic properties. The inclusion of this dataset enables assessment of the model’s robustness across a broader range of material phenomena.

To apply the TabPFN foundation model, each material is represented by either Magpie descriptors or structure-based embeddings from pre-trained graph neural network models such as ALIGNN and CGCNN. When multiple polymorphs or inconsistent values were found, we averaged the property values across similar prototypes, while outliers exceeding three standard deviations were removed.

Having established the dataset foundation, we now examine how the in-context learning foundation model (ICL-FM) performs when trained on purely compositional information versus when structural descriptors are introduced.

\subsection*{Performance of ICL-FM in Composition-Based Property Prediction}
To evaluate ICL-FM’s performance in composition based material property prediction, we benchmarked its performance on the MatBench datasets using the Magpie and MagpieEX feature representations. Table~\ref{tab:fm_magpie_full} summarizes the results and compares the FM’s mean absolute error (MAE) against the best published composition-only state-of-the-art (SOTA) models from the MatBench leaderboard~\cite{dunn2020benchmarking}. These datasets span multiple physical property classes, from exfoliation energy and phonon frequency to elastic and dielectric responses, allowing for a comprehensive assessment of the FM’s generalization capability across chemical and physical domains.

\begin{table*}[htbp]
\centering
\caption{
Comparison of composition-based performance on the MatBench datasets~\cite{dunn2020benchmarking} 
(status: 2025-08-18). 
All results are reported as mean absolute error (MAE) over the official MatBench 5-fold splits. 
Ordered by ascending dataset size, the evaluated tasks include 
\texttt{jdft2d} (meV/atom), 
\texttt{phonons} (1/cm), 
\texttt{dielectric} (unitless), 
\texttt{log\_gvrh} (log$_{10}$(GPa)), 
\texttt{log\_kvrh} (log$_{10}$(GPa)), 
and \texttt{perovskites} (meV/unit cell). 
For each property, we compare the best published MatBench composition-only SOTA with our foundation model (FM) trained on Magpie and MagpieEX feature representations, including Crabnet and Finder with composition only.
All values represent MAEs; bold entries indicate the lowest error (best performance), 
and underlined percentages denote the highest relative improvement over the SOTA.
}
\label{tab:fm_magpie_full}
\resizebox{\textwidth}{!}{
\begin{tabular}{lrlrrrr}
\toprule
\textbf{Property} & \textbf{Data Size} & \textbf{Composition SOTA} & \textbf{FM-Magpie} & \textbf{\% FM/SOTA} & \textbf{FM-MagpieEX} & \textbf{\% MagpieEX/SOTA} \\
\midrule
jdft2d (meV/atom)      &   636   & 45.6104 (Crabnet) & \textbf{44.7358} &  1.92              & 44.8311           &  1.71 \\
phonons (1/cm)     & 1,265   & 46.5751 (FinderC) & 46.4845          &  0.19              &\textbf{41.9480}   &  \underline{9.93} \\
dielectric (unitless)  & 4,764   &  0.3204 (FinderC) & \textbf{0.3066}  &  4.31              &  0.3071           &  4.15 \\
log\_gvrh (log$_{10}$(GPa))   & 10,987  &  0.0996 (FinderC) &  0.0936          &  \underline{6.02}  &  \textbf{0.0931}  &  6.57 \\
log\_kvrh (log$_{10}$(GPa))   & 10,987  &  0.0758 (Crabnet) &  0.0729          &  3.83              &  \textbf{0.0725}  &  4.33 \\
perovskites (meV/unit cell) & 18,928  &  0.4065 (Crabnet) &  0.4229          & -4.03              &  0.4240           & -4.31 \\
\bottomrule
\end{tabular}
}
\end{table*}

In Table \ref{tab:fm_magpie_full}, the 5th and 7th columns show the improvement percentages of the FMs' MAEs compared to the SOTA models. It can be found that across all six MatBench benchmarks, the foundation models (FMs) demonstrate competitive or superior performance compared. For smaller datasets such as \texttt{jdft2d} and \texttt{dielectric}, both the Magpie- and MagpieEX-based FMs achieve  consistent improvements ranging from 1.7\%  to 4.3\%, suggesting that the FMs' pretraining confers strong generalization even in data-scarce regimes. The largest relative performance gain (\underline{9.93}\%) is observed on the \texttt{phonons} task, where the MagpieEX representation substantially outperforms the baseline model, indicating that the inclusion of extended electronic descriptors in MagpieEX descriptors enhances the model’s ability to capture vibrational behavior. 

For the intermediate-scale datasets \texttt{log\_gvrh} and \texttt{log\_kvrh}, FM models with both feature sets yield stable performance improvements of roughly 4–6\% over those of existing SOTA baselines, confirming the ICL-FM’s robustness in learning the generalizable mappings between composition and elastic moduli. However, for the structurally sensitive \texttt{perovskites} dataset, FM models with both Magpie and MagpieEX underperform the composition-based SOTA, producing slightly higher MAEs (–4.03\% and –4.31\%, respectively). This result underscores that perovskite formation energies are dominated by subtle lattice distortions and symmetry variations that cannot be easily learned through composition-only descriptors.
This challenge can also be seen by the composition-based SOTA models for this task, which were ranked far behind structure-based methods on the overall \texttt{Matbench} leaderboard, reflecting the strong dependence of perovskite energetics on atomic geometry rather than purely elemental composition. For further details, please refer to Supplementary Note S3 and Table S2.
Overall, we found that FM models with the simple Magpie and MagpieEX descriptors can outperform those sophisticated expert-designed deep neural network models such as FinderC and Crabnet, indicating the strong generalization performance of FM models for material property prediction, especially for small datasets. 

To further understand the ICL capability of the ICL-FM model, we extracted the features from the final-layer activations of the learning transformer-based composition models such as CrabNet~\cite{wang2021compositionally} 
and Roost~\cite{goodall2020predicting} and evaluate the ICL prediction performance. As shown in Supplementary Tables S3 and S4, the ICL-FM model based on such advanced features only show marginal or inconsistent improvements over the baselines, suggesting that their attention-based encoders already saturate the accessible compositional information. 
While these models have strong performance themselves, their learned features contribute limited additional benefits via the foundation model’s in-context prediction.

Taken together, our results demonstrate that the ICL-FM can achieve highly competitive composition-based property prediction performance and can effectively generalize across most materials properties with small datasets. However it exhibits limitations for materials properties for which the structure-dependent effects dominate. To probe these differences in greater depth, we analyzed the phonon frequency prediction task, for which the ICL-FM yielded the most significant performance improvement.

\subsubsection{Phonon frequency dataset performance}

As shown in Table 1, for the \texttt{phonons} dataset, the ICL-FM trained on MagpieEX features achieves the highest performance gain of 9.93\% over the composition-based SOTA baseline. 
This dataset captures phonon frequencies at the $\Gamma$ point, reflecting lattice vibrational behavior that depends not only on atomic masses and bonding stiffness but also on chemical composition and bonding environment. 

\begin{figure*}[htbp]
    \centering
    \subfloat[Input Magpie feature space.]{
        \includegraphics[width=0.48\textwidth]{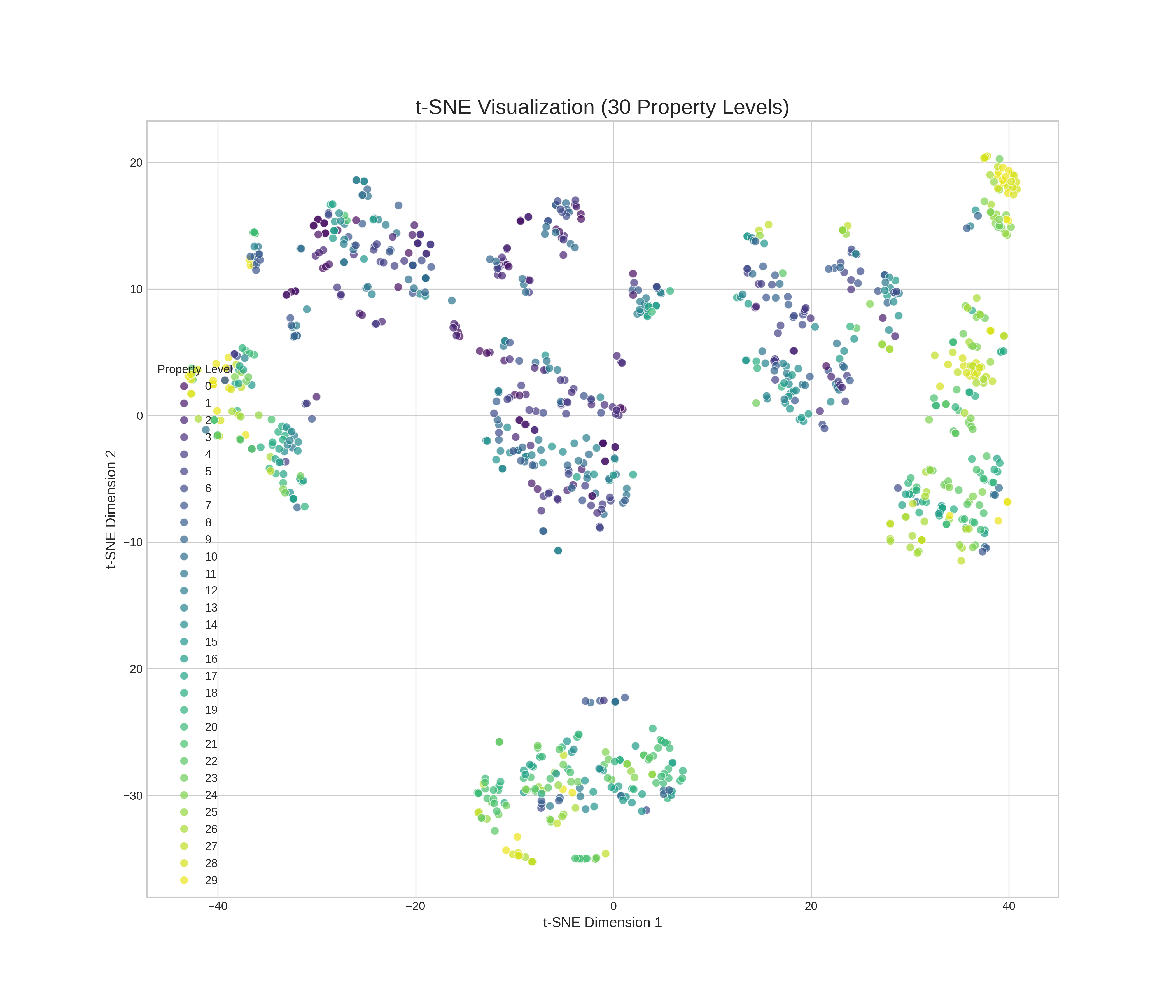}
        \label{fig:tsne_magpie_input}
    }\hspace{0.00\textwidth}
    \subfloat[FM-learned embedding space with Magpie features.]{
        \includegraphics[width=0.48\textwidth]{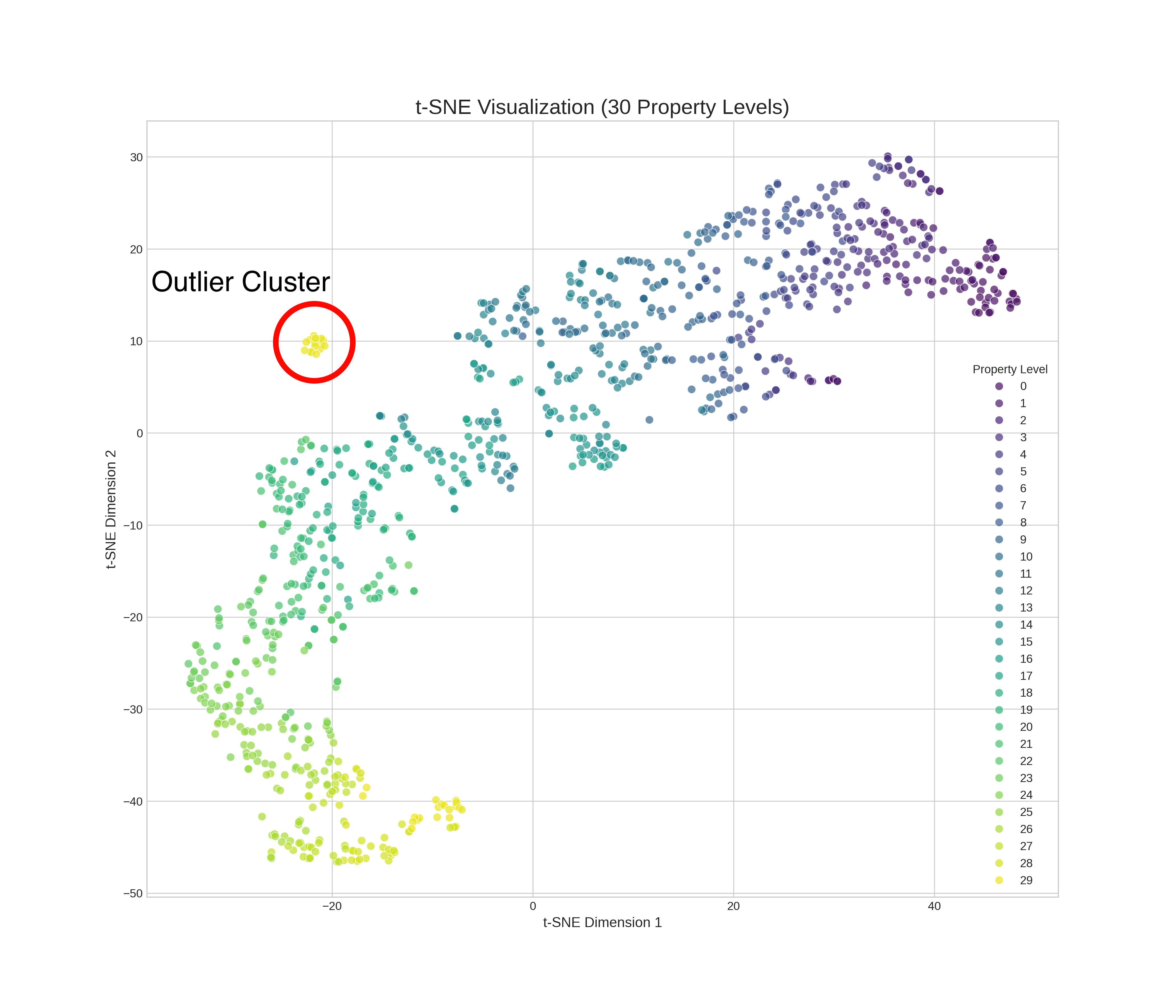}
        \label{fig:tsne_fm_magpie}
    }\\[1em]
    \subfloat[Input MagpieEX feature space.]{
        \includegraphics[width=0.48\textwidth]{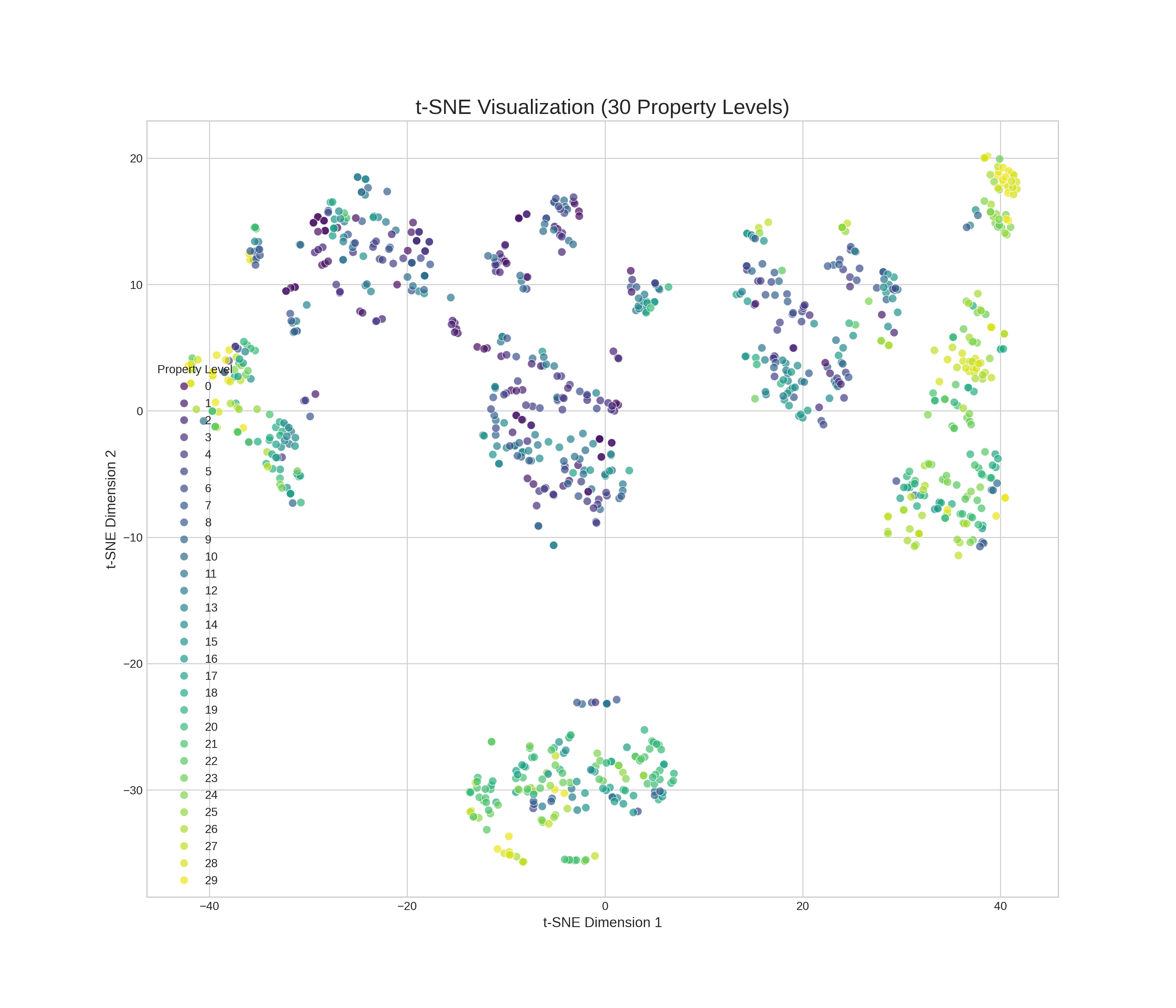}
        \label{fig:tsne_magpieEX_input}
    }\hspace{0.0\textwidth}
    \subfloat[FM-learned embedding space with MagpieEX feature.]{
        \includegraphics[width=0.48\textwidth]{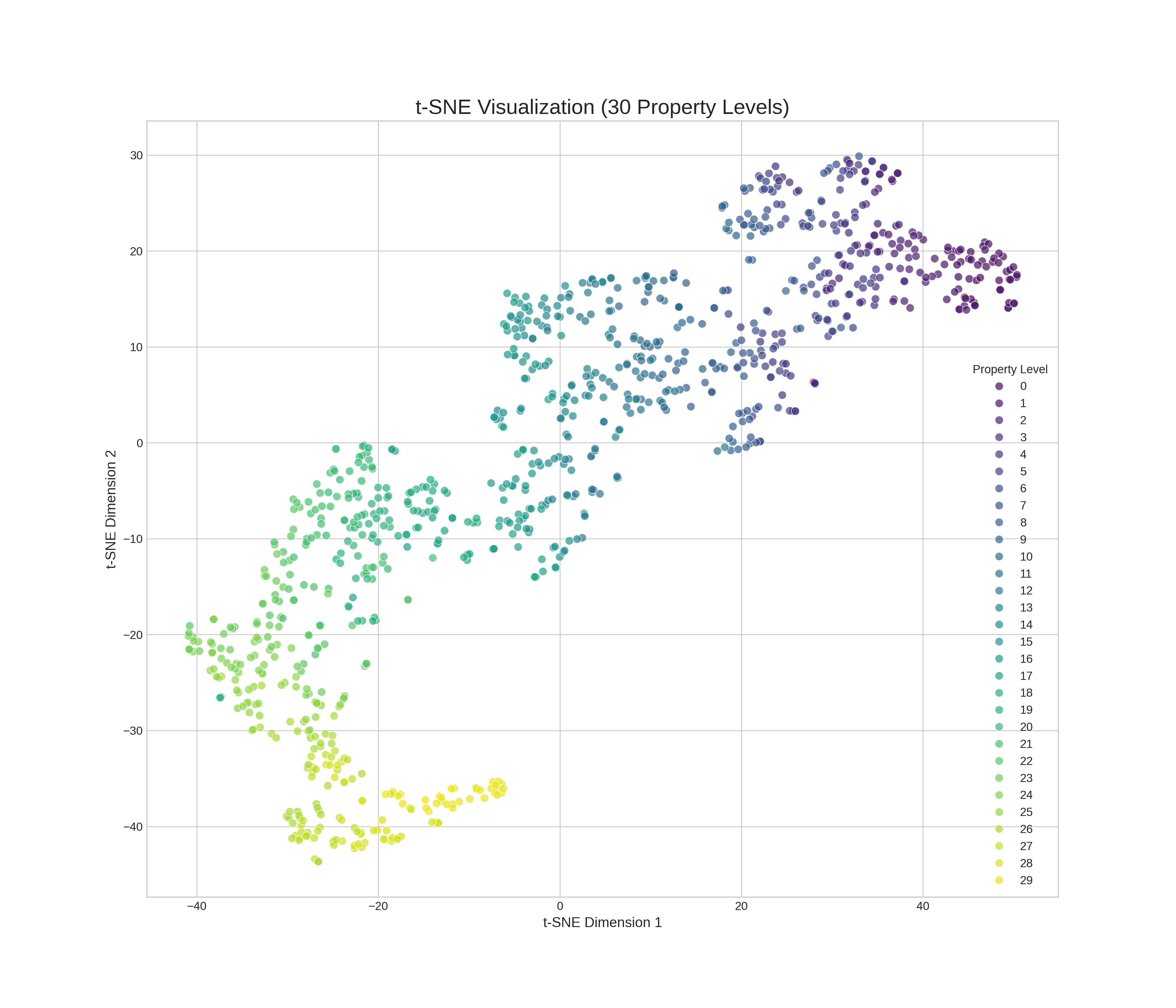}
        \label{fig:tsne_fm_magpieEX}
    }

    \caption{
    t-SNE visualization comparing the raw input feature spaces and foundation model (FM) learned embeddings for both Magpie and MagpieEX representations over phonons dataset fold 1. The colors of points indicate the material property values (phonon frequency).
    (a) The original Magpie feature space shows scattered and less distinct clusters. 
    (b) The FM embedding space learned from Magpie inputs exhibits more structured separation, indicating improved representation learning. 
    (c) The MagpieEX feature space provides a richer compositional basis. 
    (d) The FM embedding space using MagpieEX inputs produces clearer and more distinct material clusters, highlighting the benefit of extended descriptors.
    }
    \label{fig:tsne_magpie_magpieEX}
\end{figure*}

\paragraph*{t-SNE Analysis of the raw feature and learned embedding spaces }
To better understand how the ICL-FM leverages compositional information to achieve significantly improved prediction accuracy on the \texttt{phonons} dataset, we visualize the learned feature representations by ICL-FM using t-distributed stochastic neighbor embedding (t-SNE) \cite{maaten2008visualizing}. 
Figure~\ref{fig:tsne_magpie_magpieEX} compares the input feature spaces and the FM-learned embeddings for both Magpie and MagpieEX representations. 
Here we aim to reveal how the foundation model reorganizes the high-dimensional feature space into more physically meaningful manifolds that correlate with phonon frequency levels, providing insight into the source of the observed 9.93\% performance gain.

In Figure~\ref{fig:tsne_magpie_magpieEX}, the transition from panels (a)\&(b) to panels (c)\&(d) illustrates a progressive transformation of the feature space by the ICL-FM model. 
In the original Magpie feature space (a), data points are dispersed into irregular clusters, whereas after FM processing (b), the embeddings exhibit smoother and more continuous organization in both the continuity of the sample distribution and the continuity and gradual progressing of the property values, both leading to higher generalization and prediction performance. 
This explains why the FM’s in-context learning (ICL) mechanism enhances predictive performance. 
Notably, the isolated yellow cluster observed in panel (b) corresponds to outlier materials with extreme phonon frequencies, suggesting that the FM-ICL can identify such rare regimes even with the limited compositional information. This well-behaviored property value distribution can bring high potential for discovering materials with extreme properties. 

Figure~\ref{fig:tsne_magpie_magpieEX} (d) shows the FM-learned feature space using MagpieEX inputs. We found that these outlier points merge more seamlessly with the main cluster, showing a smoother color transition that corresponds to continuous variation in phonon frequency. 
This demonstrates that the combination of extended compositional descriptors and the FM’s learned embedding space yields more physically coherent representations.

\begin{figure*}[htbp]
    \centering
    \subfloat[SHAP feature importance for Magpie features.]{
        \includegraphics[width=0.48\textwidth]{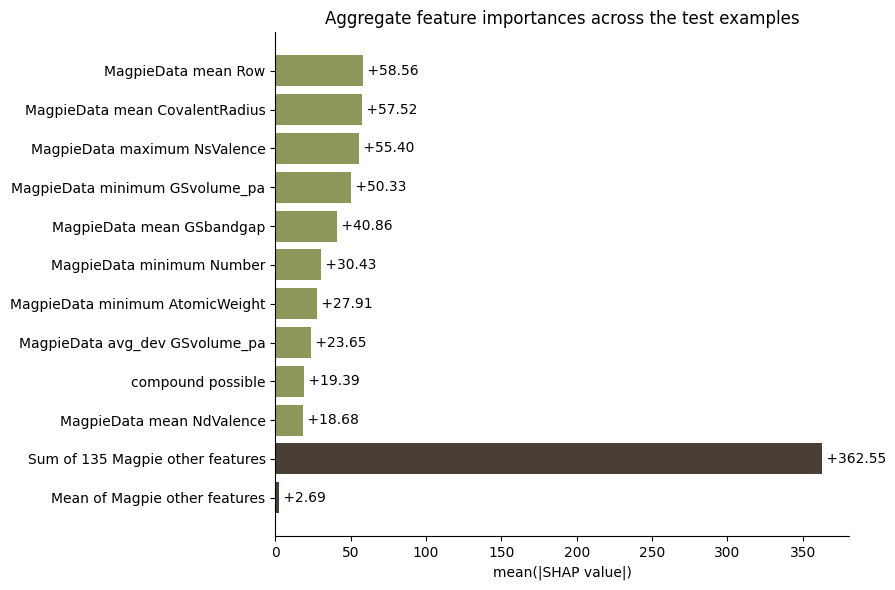}
        \label{fig:shap_magpie}
    }\hfill
    \subfloat[SHAP feature importance for MagpieEX features.]{
        \includegraphics[width=0.48\textwidth]{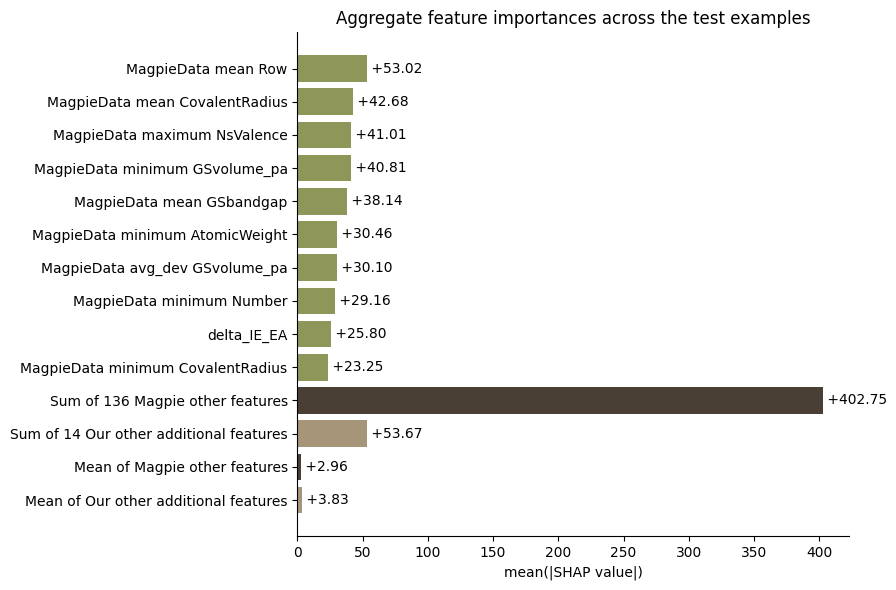}
        \label{fig:shap_magpieEX}
    }

    \caption{
    Comparison of SHAP analyses for Magpie and MagpieEX feature inputs of their aggregate feature importance.  
    (a) SHAP summary plot showing the contribution of individual Magpie features to the FM’s predictions.  
    (b) SHAP summary plot for MagpieEX features, illustrating the additional influence of extended compositional descriptors on model interpretability.
    }
    \label{fig:shap_bar}
\end{figure*}

\paragraph*{SHAP Feature Importance Analysis}
While the t-SNE visualization reveals how the FM-ICL transforms the global feature structure, it is interesting to achieve a more detailed understanding of its decision process, by linking model predictions to feature-level contributions.
Here we use explanatory SHAP (SHapley Additive exPlanations) analysis to quantify how individual compositional features contribute to the ICL-FM’s predictions on the \texttt{phonons} frequency.  Figure~\ref{fig:shap_bar} compares the aggregate feature importance between the Magpie and MagpieEX features, highlighting how extended descriptors alter the model’s interpretation of chemical information.

The SHAP analysis results in Figure~\ref{fig:shap_bar}(a) identify several dominant compositional descriptors in the Magpie representation, including mean periodic table row, mean covalent radius, and maximum number of $s$-valence electrons.  
These features indicate that the model relies primarily on atomic size, elemental periodicity, and valence configuration—key chemical factors linked to bonding strength and lattice rigidity that strongly affect phonon vibrations.  
Additional descriptors such as minimum ground-state volume per atom and mean ground-state band gap reveal sensitivity to structural compactness and electronic insulation tendencies, suggesting that even without geometric input, the ICL-FM can infer approximate bonding and stability trends from compositional statistics.

In contrast, the MagpieEX feature ranking (Figure~\ref{fig:shap_bar}(b)) introduces extended electronic-level descriptors such as $\Delta(\mathrm{IE}-\mathrm{EA})$, the difference between cation ionization energy and anion electron affinity, which serves as a proxy for bond ionicity and charge-transfer strength, leading to its better phonon frequency prediction performance.  
The emergence of this feature among the top SHAP features shows that the ICL-FM benefits from descriptors explicitly encoding electronic polarization, enabling richer and more physically grounded interpretation.  
Together, these results confirm that the ICL-FM with MagpieEX representation improves both predictive accuracy and interpretability by embedding chemical bonding information directly into the compositional feature space. 
Detailed SHAP and feature-dependence analyses that illustrate per-sample contributions and feature interactions are provided in Supplementary Figure S2.

Our results demonstrate that the in-context learning (ICL) capability of the foundation model can deliver competitive or even superior performance compared to advanced composition-based architectures, despite relying solely on rudimentary compositional descriptors such as the raw Magpie inputs. 
Remarkably, the foundation model achieves these results \emph{without any supervised fine-tuning or parameter updates}, drawing exclusively on its pretrained reasoning and few-shot inference-time adaptation capacity. 
This finding underscores that the FM’s internal contextualization mechanism effectively encodes compositional relationships that traditional deep learning models typically acquire only through extensive, task-specific training.

The observations highlight a key strength of foundation-driven ICL-FM approaches: 
they achieve strong generalization from minimal input complexity and training effort, revealing that compositional reasoning can emerge directly from simple tabular descriptors without reliance on elaborate neural network architectures or extensive retraining. 
However, composition-only representations inherently omit the geometric and topological factors that govern most structure-sensitive properties such as formation energy or mechanical stability. To address this limitation, the next section extends our investigation to \emph{structure-based property prediction}, where graph-derived embeddings from ALIGNN and CGCNN are integrated with the FM.

\subsection*{ICL-FM for Structure-Based Property Prediction}

\subsubsection*{GNN Models: General performance improvement but below SOTA results}
Building on the success of ICL-FM in composition-based property prediction, we assess the effectiveness of ICL-FM in structure based  property prediction tasks. 
We applied our ICL-FM framework to graph neural network (GNN) architectures using the MatBench benchmark datasets. We use the GNN models pretrained for material property prediction to extract advanced features, which are then fed to the ICL-FM models for learning-free property prediction. We call such models as FMEF (\emph{Foundation Models with Extracted Features}).
The training details of these GNN models are shown in Supplementary Table S1.

Table~\ref{tab:farm_alignn_matbench} summarizes the results comparing the proposed \textbf{FMEF} against the ALIGNN baseline, a leading GNN model for material property prediction and other SOTA models. We include the results of both the original ALIGNN benchmark~\cite{choudhary2021alignn} and a locally modified ALIGNN baseline implementation. 
In these experimental settings, FMEF models essentially is used to replace the baseline’s final fully connected layer after its supervised training with a pretrained TabPFN module, allowing the model to leverage prior tabular reasoning for feature-to-property mapping.
We report mean absolute errors (MAEs) across the standard 5-fold splits, with relative performance improvements computed against both the benchmark ALIGNN and the local ALIGNN baseline.

\begin{table}[!htp]\centering

\caption{
Performance comparison of FMEF with ALIGNN and SOTA models on MatBench structure-based property prediction datasets and splits 
\cite{dunn2020benchmarking} (status 2025-08-18). 
All results are reported as mean absolute errors (MAEs) over the official MatBench 5-fold splits. 
Current benchmark holders include MODNet \cite{debreuck2021modnet}, 
MegNet \cite{chen2019graph}, and coNGN \cite{ruff2023connectivity}. 
We additionally report ALIGNN from the official leaderboard \cite{choudhary2021alignn}, 
our modified ALIGNN \emph{baseline} implementation (adapted for compatibility with foundation models), and our proposed \textbf{FMEF}, 
which replaces the baseline’s final fully connected layer with TabPFN. 
Relative improvements (\%) are shown with respect to both the ALIGNN benchmark 
and the local ALIGNN baseline. The best MAE per dataset is highlighted in bold.
}
\label{tab:farm_alignn_matbench}
\resizebox{\textwidth}{!}{ %
\begin{tabular}{lrrrrrrr}
\toprule
Property   &Data Size &SOTA& ALIGNN benchmark&Baseline &FMEF &\%Impr. benchmark &\% Impr. baseline \\\midrule
jdft2d (meV/atom)    &636     &33.1918 (MODNet) & 43.4244 &41.1251 &\textbf{40.8797} &5.85 &0.59 \\
phonons (1/cm)    &1,265   &28.7606 (MegNet) & 29.5385 &\textbf{24.9136} &29.7533 &-0.73 &-19.43 \\
dielectric (unitless) &4,764   &0.2711 (MODNet)  & 0.3449  &0.2866 &\textbf{0.2743} &\textbf{20.47} &4.30 \\
log\_gvrh (log$_{10}$(GPa))  &10,987  &0.0670 (coNGN)   & 0.0715  &0.0703 &\textbf{0.0702} &1.76 &0.11 \\
log\_kvrh (log$_{10}$(GPa))  &10,987  &0.0491 (coNGN)   & 0.0568  &0.0539 &\textbf{0.0533} &6.11 &1.07 \\
perovskites (meV/unit cell) &18,928  &0.0269 (coGN)    & 0.0288  &0.0303 &0.0318 &-10.44 &-4.82 \\
\bottomrule
\end{tabular}
} %
\end{table}

First from Table 2, we found that across the six structure-based MatBench tasks, the proposed \textbf{FMEF} framework achieves consistent or modest gains relative to the ALIGNN benchmark, demonstrating that integrating pretrained TabPFN model with extracted features from GNN models can enhance GNN performance for structure-informed property prediction. However, none of the FMEF results outperform the SOTA (column 3), different from the composition-based FM models. 
Notable relative improvements are observed for the \texttt{dielectric} (+20.47\%) and \texttt{jdft2d} (+5.85\%) datasets, where the extracted latent features likely facilitate a more effective mapping between structural and electronic attributes to the exfoliation energy and refractive index properties.
For intermediate-scale datasets such as \texttt{log\_gvrh} and \texttt{log\_kvrh}, the performance gains of FMEF models are smaller (1–6\%) but stable, suggesting that the benefit of pretrained priors decreases as the data volume becomes sufficient for direct supervised training.

Interestingly, our locally tuned ALIGNN baseline outperforms the published SOTA result for the \texttt{phonons} dataset (MAE = 24.91 vs.\ 28.76 by MegNet), indicating that our revised training setup and hyperparameter calibration substantially improve model efficiency and generalization. 
However, the FMEF model based on features extracted from this ALIGNN baseline shows a performance decline in this case, likely because replacing the final dense layer with TabPFN introduces over-regularization or reduces the model’s flexibility to capture subtle vibrational interactions. 
A similar effect occurs for the \texttt{perovskites} dataset, where strong structure–property coupling 
limits the effectiveness of FM-based tabular adaptation.

To further validate the generality of this integration strategy, we evaluated the FMEF framework across multiple additional neural network models, including structure-based models such as CGCNN~\cite{xie2018crystal}, coNGN~\cite{ruff2023connectivity}, and DeeperGATGNN~\cite{omee2022scalable}. 
For each model, we applied the same foundation adaptation protocol and consistent hyperparameter settings to ensure compatibility between the pretrained TabPFN component and the native network encoder. 
The complete results are provided in the Supplementary Note S2 and Tables S5–S7. It was found that no ICL-FM models based on features extracted from such baseline GNN models can outperform the SOTA GNN models on such structure-based property prediction tasks, showing the limited capability of FM models for learning complex structure-property relationship using in-context learning only. 
In addition, the pre-trained CGCNN model checkpoint is used to evaluate the transfer learning capacity of the ICL-FM. The detailed results are presented in Supplementary Note S4 and Table S8. 

Overall, these results show that the FMEF architecture can improve graph-based prediction performance for some materials properties with appropriate structure-based GNN models. However, it may also fail for properties dominated by complex atomic interactions. 
Compared to composition-based property prediction, FM performance improvement is dependent on the dataset, the baseline model, and the material property, showing a much more complex scenario.
This emphasizes that, while foundation model integration in general can provide promising generalization benefits, future designs should tailor the coupling strategy between pretrained reasoning modules and task-specific encoders to the physical complexity of the target property.

\subsubsection*{Layer-Wise Feature Contribution for Property Prediction via Foundation Model Adaptation}
Another interesting usage of ICL-FM is to investigate how information evolves across successive network layers of a GNN by quantifying 
the contribution of each layer’s learned representation to downstream performance. 
Rather than removing layers through ablation, we use the foundation model (FM) as a diagnostic probe to evaluate the representation power of intermediate embeddings extracted from different network depths.
Specifically, we first extract intermediate embeddings from each ALIGNN layer, covering both ALIGNN message-passing blocks (\texttt{ali0}–\texttt{ali3}) and subsequent GCN refinement layers (\texttt{gcn0}–\texttt{gcn3}), as well as the final graph-level \texttt{readout} representation. 
Each set of these features is then independently passed into the FM for inference-time property prediction.  
The mean absolute errors (MAEs) over five folds are summarized in Table~\ref{tab:layer_fm_results}, highlighting which intermediate embeddings carry the most informative structural information for the FM to leverage.

\begin{table}[htbp]
\centering
\caption{
Layer-wise foundation model (FM) performance on the dielectric dataset.
Embeddings are extracted from different stages of the ALIGNN architecture. 
The network consists of four ALIGNN layers (\texttt{ali0–ali3}), four GCN layers (\texttt{gcn0–gcn3}), 
Reported values correspond to the mean absolute error (MAE) of refractive index prediction (unitless), averaged over five folds; lower MAE indicates better performance 
after FM adaptation.
}
\label{tab:layer_fm_results}
\resizebox{0.9\textwidth}{!}{
\begin{tabular}{lccccccccc}
\toprule
& \multicolumn{4}{c}{\textbf{ALIGNN layers}} 
& \multicolumn{4}{c}{\textbf{GCN layers}} 
& \multicolumn{1}{c}{\textbf{Readout}} \\
\cmidrule(lr){2-5} \cmidrule(lr){6-9} \cmidrule(lr){10-10}
\textbf{Layer index} 
& \texttt{ali0} & \texttt{ali1} & \texttt{ali2} & \texttt{ali3} 
& \texttt{gcn0} & \texttt{gcn1} & \texttt{gcn2} & \texttt{gcn3} 
& \texttt{readout} \\
\midrule
\textbf{MAE (5-fold mean)} 
& 0.3088 & 0.3065 & 0.2952 & \textbf{0.2921} 
& 0.3130 & 0.3149 & 0.3141 & \textbf{0.3118} 
& \textbf{0.2894} \\
\bottomrule
\end{tabular}
}
\end{table}

As shown in Table~\ref{tab:layer_fm_results}, on the dielectric dataset, the FM performance improves as representations progress from shallow (\texttt{ali0}) layers to deeper (\texttt{gcn} and \texttt{readout}) layers. 
The lowest MAEs are achieved for the features extracted from the final ALIGNN layer (\texttt{ali3}) and the graph-level \texttt{readout} representation (0.2894), indicating that deeper embeddings encode more complete atomic–bond–graph correlations that the FM can most effectively leverage. This layer-wise progression reflects how deeper GNN layers capture increasingly holistic structural and chemical information, enhancing the FM’s ability to perform accurate feature-to-property mapping.
Therefore, instead of relying on iterative retraining or pruning, 
the FM-based adaptation approach provides a systematic and efficient way to identify the most transferable representation layers in hierarchical GNN architectures.

\subsubsection*{Feature fusion for ICL-FM based property prediction}

A natural extension of above feature-level study is to check whether integrating features from different levels can improve the ICL-FM performance, similar to the U-Net network \cite{ronneberger2015u} which combines multi-resolution features for high-performance prediction. 
Here, we combine outputs from multiple layers of a GNN and feed them to the TabPFN network for prediction.
Specifically, we evaluate three feature integration strategies and compare to the baseline ALIGNN model:

\begin{itemize}
    \item[(1)] \textbf{GNN Input + FM}: the FM operates directly on the raw atomic and bonding features.
    \item[(2)] \textbf{Latent Feature + FM}: the FM is trained on the final-layer node embeddings extracted from ALIGNN.
    \item[(3)] \textbf{Feature Fusion + FM}: multi-layer ALIGNN features are concatenated before FM adaptation.
\end{itemize}

\begin{table}[htbp]
\centering
\caption{
Performance comparison of different ALIGNN–FM feature integration strategies on MatBench structure datasets~\cite{dunn2020benchmarking}
(status: 2025-08-18). 
All results are reported as mean absolute errors (MAEs) over the official five-fold splits. 
\textbf{GNN Input + FM} uses raw graph inputs, 
\textbf{Latent Feature + FM} applies the FM to the final readout embeddings from ALIGNN, 
and \textbf{Feature Fusion + FM} concatenates multi-layer ALIGNN features before FM adaptation. 
The best MAE of each dataset is highlighted in bold.
}
\label{tab:alignn_fm_comparison}
\resizebox{\textwidth}{!}{
\begin{tabular}{lrrrrrr}
\toprule
\textbf{Property} & \textbf{Size} & 
\textbf{GNN Input + FM} & \textbf{ALIGNN} & 
\textbf{Latent Feature + FM} & \textbf{Feature Fusion + FM} \\
\midrule
\texttt{jdft2d} (meV/atom)      & 636     & \textbf{39.4571} & 41.1251 & 40.8797 & 40.9175 \\
\texttt{phonons} (1/cm)    & 1,265   & 54.1106 & \textbf{24.9136} & 29.7533 & 29.9737 \\
\texttt{dielectric} (unitless)  & 4,764   & 0.5192  & 0.2866 & \textbf{0.2743} & 0.2788 \\
\texttt{log\_gvrh} (log$_{10}$(GPa))   & 10,987  & 0.1293  & 0.0703 & \textbf{0.0702} & \textbf{0.0702} \\
\texttt{log\_kvrh} (log$_{10}$(GPa))  & 10,987  & 0.1073  & 0.0539 & \textbf{0.0533} & 0.0534 \\
\texttt{perovskites} (meV/unit cell) & 18,928  & 0.0781  & \textbf{0.0303} & 0.0318 & 0.0327 \\
\bottomrule
\end{tabular}
}
\end{table}

The results on six MatBench structure-based datasets are summarized in Table~\ref{tab:alignn_fm_comparison}. The three kinds of feature inputs to FM are described in the Method section.
Across all datasets, directly feeding raw ALIGNN inputs into the FM leads to degraded performance, indicating that the FM alone cannot model complex correlations between structures and material properties without prior graph encoding. In contrast, coupling the FM with higher-level ALIGNN representations, particularly the final-layer embeddings, consistently improves accuracy over the standalone ALIGNN baseline. 
The \textbf{Latent Feature + FM} configuration achieves the best or near-best MAE in most cases (e.g., \texttt{dielectric}, \texttt{log\_gvrh}, and \texttt{log\_kvrh}), confirming that the FM benefits most from semantically rich, structure-aware embeddings generated by the GNN encoder. The \textbf{Feature Fusion + FM} strategy produces similar but slightly less consistent results, suggesting that concatenating multi-layer features offers limited additional advantage once the FM has access to the fully aggregated latent representation.

These results demonstrate that the foundation model can not only function as an analytical tool for understanding GNN representation hierarchies but also can serve as a lightweight, performance-enhancing component for material property prediction with small datasets. By integrating the FM with features extracted from different stages of the GNN, we systematically uncovered how structural and compositional information contributes to predictive performance, offering a scalable, training-free alternative to conventional ablation-based interpretability methods.

\subsection*{Lattice Thermal Conductivity (LTC) prediction using ICL-FM}

To further assess the generality of the proposed in-context learning foundation model (ICL-FM) framework, we extend our analysis beyond the MatBench benchmark suite to the prediction of lattice thermal conductivity (LTC), a practically important material macro-property.  
The LTC property provides a stringent test of model generalization, as it depends sensitively on anharmonic lattice vibrations, phonon–phonon scattering, and atomic mass contrast, phenomena that require an integrated understanding of both compositional and structural effects. 

\paragraph*{LTC prediction performance using Magpie-based compositional and structural descriptors}
We first evaluate the ICL-FM on composition- and structure-augmented tabular descriptors to examine its adaptability to heterogeneous handcrafted features. 
The experiments use four progressively enriched feature sets Magpie, MagpieEX, Magpie+Struct1, and Magpie+Struct2 that capture compositional, bonding, and geometric characteristics. All composition and structure features are calculated using the matminer package. These feature sets are defined below:

\begin{itemize}
    \item \textbf{Magpie}: A composition-only feature set that combines 
    stoichiometric statistics, elemental property averages (Magpie preset), 
    valence orbital occupancies, and ionization properties. 
    It serves as the baseline representation capturing average compositional trends.

    \item \textbf{MagpieEX}: Extends Magpie with cation–anion contrast descriptors, 
    including differences in electronegativity, ionic radius, valence orbital populations, 
    ionization potential minus electron affinity, and polarizability, 
    as well as derived quantities such as bond ionicity, oxidation asymmetry, and lone-pair indicators. 
    These features enhance the representation of bonding polarity and charge-transfer effects.

    \item \textbf{Magpie+Struct1}: Incorporates additional global structural descriptors, 
    including density and symmetry statistics, to capture geometric influences 
    that complement the compositional features. 
    This representation bridges atomic composition and crystal-level organization.

    \item \textbf{Magpie+Struct2}: Combines Magpie+Struct1 with richer geometric and topological descriptors, 
    such as structural heterogeneity, minimum relative distances, and radial distribution functions, 
    providing a more complete description of atomic-scale packing and local coordination environments.
\end{itemize}

For comparison, conventional Random Forest (RF) and multilayer perceptron (MLP) models are also trained and evaluated using identical five-fold cross-validation. 
Both mean absolute errors (MAEs) and coefficient of determination ($R^2$) are reported, and the results are summarized in Table~\ref{tab:fm_thermo_compo}.

\begin{table}[htbp]
\centering
\caption{
Performance comparison on the standalone lattice thermal conductivity dataset using different feature representations 
(Magpie, MagpieEX, Magpie+Struct1, and Magpie+Struct2) and models 
(Foundation Model (FM), Random Forest (RF), and MLP). 
Values represent the mean $\pm$ standard deviation of the mean absolute error (MAE, in ($\mathrm{W\,m^{-1}\,K^{-1}}$)) and the coefficient of determination ($R^2$), averaged over five folds.
The best MAE (lowest) and best $R^2$ (highest) per feature set are highlighted in bold.
}
\label{tab:fm_thermo_compo}
\resizebox{\textwidth}{!}{
\begin{tabular}{lcccccc}
\toprule
\textbf{Feature Set} & 
\multicolumn{2}{c}{\textbf{FM}} & 
\multicolumn{2}{c}{\textbf{Random Forest (RF)}} & 
\multicolumn{2}{c}{\textbf{MLP}} \\
\cmidrule(lr){2-3} \cmidrule(lr){4-5} \cmidrule(lr){6-7}
 & MAE ($\mathrm{W\,m^{-1}\,K^{-1}}$) ($\downarrow$) & $R^2$ ($\uparrow$) & MAE ($\mathrm{W\,m^{-1}\,K^{-1}}$) ($\downarrow$) & $R^2$ ($\uparrow$) & MAE ($\mathrm{W\,m^{-1}\,K^{-1}}$) ($\downarrow$) & $R^2$ ($\uparrow$) \\
\midrule
\textbf{Magpie}          & \textbf{7.59 $\pm$ 2.89} & \textbf{0.34 $\pm$ 0.15} & 9.90 $\pm$ 2.50 & 0.04 $\pm$ 0.41 & 14.04 $\pm$ 3.87 & 0.07 $\pm$ 0.29 \\
\textbf{MagpieEX}        & \textbf{7.63 $\pm$ 2.91} & \textbf{0.34 $\pm$ 0.15} & 9.95 $\pm$ 2.49 & 0.03 $\pm$ 0.41 & 12.03 $\pm$ 3.25 & 0.20 $\pm$ 0.13 \\
\textbf{Magpie+Struct1}         & \textbf{7.41 $\pm$ 2.92} & \textbf{0.36 $\pm$ 0.16} & 9.61 $\pm$ 2.33 & $-$0.06 $\pm$ 0.59 & 13.60 $\pm$ 2.99 & 0.06 $\pm$ 0.28 \\
\textbf{Magpie+Struct2}   & \textbf{7.45 $\pm$ 2.90} & \textbf{0.37 $\pm$ 0.17} & 9.99 $\pm$ 2.43 & $-$0.23 $\pm$ 0.85 & 12.65 $\pm$ 2.97 & $-$0.83 $\pm$ 2.01 \\
\bottomrule
\end{tabular}
}
\end{table}

As shown in Table~\ref{tab:fm_thermo_compo}, the foundation model (FM) consistently achieves 
the best predictive performance across all four feature sets, 
outperforming conventional Random Forest (RF) and multi-layer perceptron (MLP) baselines in both MAE and $R^2$. 
Across five-fold evaluations, the FM achieves an average MAE of approximately 7.5–7.6, 
with $R^2$ values exceeding 0.33 for all compositional and structure-augmented representations. 
In contrast, RF and MLP models exhibit higher prediction errors 
(MAE in the range of 9-14) and unstable $R^2$ values, including negative correlations 
for structure-rich feature spaces. 
This highlights the FM’s robust capacity to generalize from limited data 
and to capture higher-order nonlinear interactions among features 
that conventional models struggle to learn.

Among the feature types, the inclusion of extended or structural descriptors 
in MagpieEX, Magpie+Struct1, and Magpie+Struct2 yields incremental improvements 
over the base Magpie features, underscoring the importance of structural 
and bonding information in thermal transport modeling. 
MagpieEX enriches the feature space with charge-transfer and ionicity descriptors, 
allowing the FM to better capture bonding asymmetry and electronic screening effects, factors 
closely tied to phonon scattering and lattice anharmonicity. 
The Magpie+Struct1 and Magpie+Struct2 feature sets further enhance predictive accuracy 
by incorporating explicit structural attributes such as density, 
symmetry, and local coordination statistics, which influence phonon mean free paths 
and heat-carrying modes. 
These incremental improvements suggest that while composition alone 
captures broad chemical trends, structural information is critical 
for resolving fine-grained variations in thermal transport behavior.

Notably, the FM demonstrates remarkable consistency across all feature categories, 
achieving similar MAE and $R^2$ performance even as the input dimensionality 
and physical abstraction level increase. 
This indicates that the FM’s in-context learning mechanism adapts effectively 
to both simple compositional and high-dimensional structural representations, 
providing stable generalization without extensive hyperparameter tuning. 
By comparison, the RF and MLP models show larger variance and occasional overfitting, 
especially for structure-heavy descriptors, reflecting their limited ability 
to balance complex feature correlations with small-sample datasets.

These results confirm that the proposed FM framework 
can extract transferable and physically meaningful representations 
from heterogeneous input spaces, enabling accurate and data-efficient prediction 
of lattice thermal conductivity. 
The observed performance hierarchy, Magpie $\rightarrow$ MagpieEX $\rightarrow$ Magpie+Struct1 
$\rightarrow$ Magpie+Struct2, further illustrates that combining compositional, 
bonding, and structural priors leads to a progressively richer and more 
physics-grounded feature manifold for modeling phonon-mediated heat transport. 
This suggests strong potential for applying the FM approach to other 
structure-sensitive material properties, such as carrier mobility, 
thermal expansion, and elastic anisotropy.

\paragraph*{LTC Prediction performance using ALIGNN graph embeddings}
We next examine the effectiveness of the ICL-FM on structure-aware representations 
by integrating it with the ALIGNN architecture for predicting lattice thermal conductivity. 
Two FM coupling strategies are considered:  
(1) \textbf{ALIGNN–FM (last-layer features)}, in which the FM is trained on the final ALIGNN embeddings, and  
(2) \textbf{Feature-Fusion FM}, where multi-layer ALIGNN features are concatenated before FM adaptation.  
The baseline ALIGNN model without FM coupling is included for comparison.  
Table~\ref{tab:alignn_fm_thermo} summarizes the mean absolute error (MAE), 
coefficient of determination ($R^2$), 
and Spearman rank correlation coefficient, which evaluates the monotonic consistency between predicted and true thermal conductivity values, 
and relative improvement over the baseline across five folds.

\begin{table}[htbp]
\centering
\caption{
Performance comparison of ICL-FMs with two feature integration strategies on the standalone thermal conductivity dataset. 
The \textbf{ALIGNN} model serves as the baseline, 
while \textbf{ALIGNN--FM} uses last-layer graph embeddings as input to the FM, 
and \textbf{Feature-Fusion FM} leverages concatenated multi-layer embeddings. 
Values represent the mean MAE, mean $R^2$, and mean \textbf{Spearman rank correlation} averaged over five folds. 
Relative improvement (\%) is computed with respect to the baseline MAE. 
Spearman correlation is reported as it better captures monotonic relationships and is more suitable for material property prediction tasks.
}
\label{tab:alignn_fm_thermo}
\resizebox{0.95\textwidth}{!}{
\begin{tabular}{lcccc}
\toprule
\textbf{Model} 
& \textbf{mean MAE ($\mathrm{W\,m^{-1}\,K^{-1}}$) ($\downarrow$)} 
& \textbf{mean $R^2$ ($\uparrow$)} 
& \textbf{mean Spearman ($\uparrow$)} 
& \textbf{MAE Impv.(\%) ($\uparrow$)} \\
\midrule
\textbf{ALIGNN (baseline)} 
& 8.972 
& 0.2339 
& $0.8525 \pm 0.0136$ 
& -- \\

\textbf{ALIGNN--FM (last-layer)} 
& 8.701 
& 0.2682 
& $0.8589 \pm 0.0032$ 
& \textbf{+3.02} \\

\textbf{Feature-Fusion FM} 
& \textbf{8.672} 
& 0.2664 
& \textbf{$0.8631 \pm 0.0058$} 
& \textbf{+3.34} \\
\bottomrule
\end{tabular}
}
\end{table}

Both FM-enhanced configurations outperform the standalone ALIGNN model, 
achieving consistent reductions in MAEs and modest improvements in $R^2$. 
The \textbf{ALIGNN–FM} model (last-layer features) reduces the mean MAE from 8.97 to 8.70 
(approximately 3.0\% improvement), while the \textbf{Feature-Fusion FM} configuration 
further decreases the MAE to 8.67 (3.3\% improvement). 
In addition to lower prediction error, both FM-based variants exhibit improved Spearman ranking correlations, indicating better preservation of the relative LTC ordering of materials.
This incremental gain suggests that combining embeddings from multiple ALIGNN layers 
enables the FM to leverage both low-level atomic connectivity and higher-order graph context, 
resulting in smoother, more physically meaningful representations of structural information.  

Although the absolute performance gains appear moderate, 
they are significant given the strong baseline of the ALIGNN architecture and the intrinsic difficulty of lattice thermal conductivity prediction. 
The consistent improvement in Spearman ranking correlation coefficient (Spearmanr) further highlights that FM adaptation enhances ranking robustness, which is particularly important for materials screening and discovery tasks where relative trends often matter more than exact values.
These findings confirm that FMs offer a lightweight yet effective performance improvement of the GNN models by improving feature generalization without additional structural training or model retraining.

\subsubsection*{Interpretability and Physical Insights into Thermal Conductivity}
To elucidate the physical mechanisms governing lattice thermal conductivity ($\kappa$), we analyze feature importance and SHAP values from both the ICL-FM and the Random Forest (RF) baseline. 
This comparison reveals how each model encodes structural stiffness, compositional disorder, and electronic character when predicting thermal transport.

\begin{figure}[htbp]
    \centering
    \begin{minipage}[b]{0.49\textwidth}
        \centering
        \includegraphics[width=\textwidth]{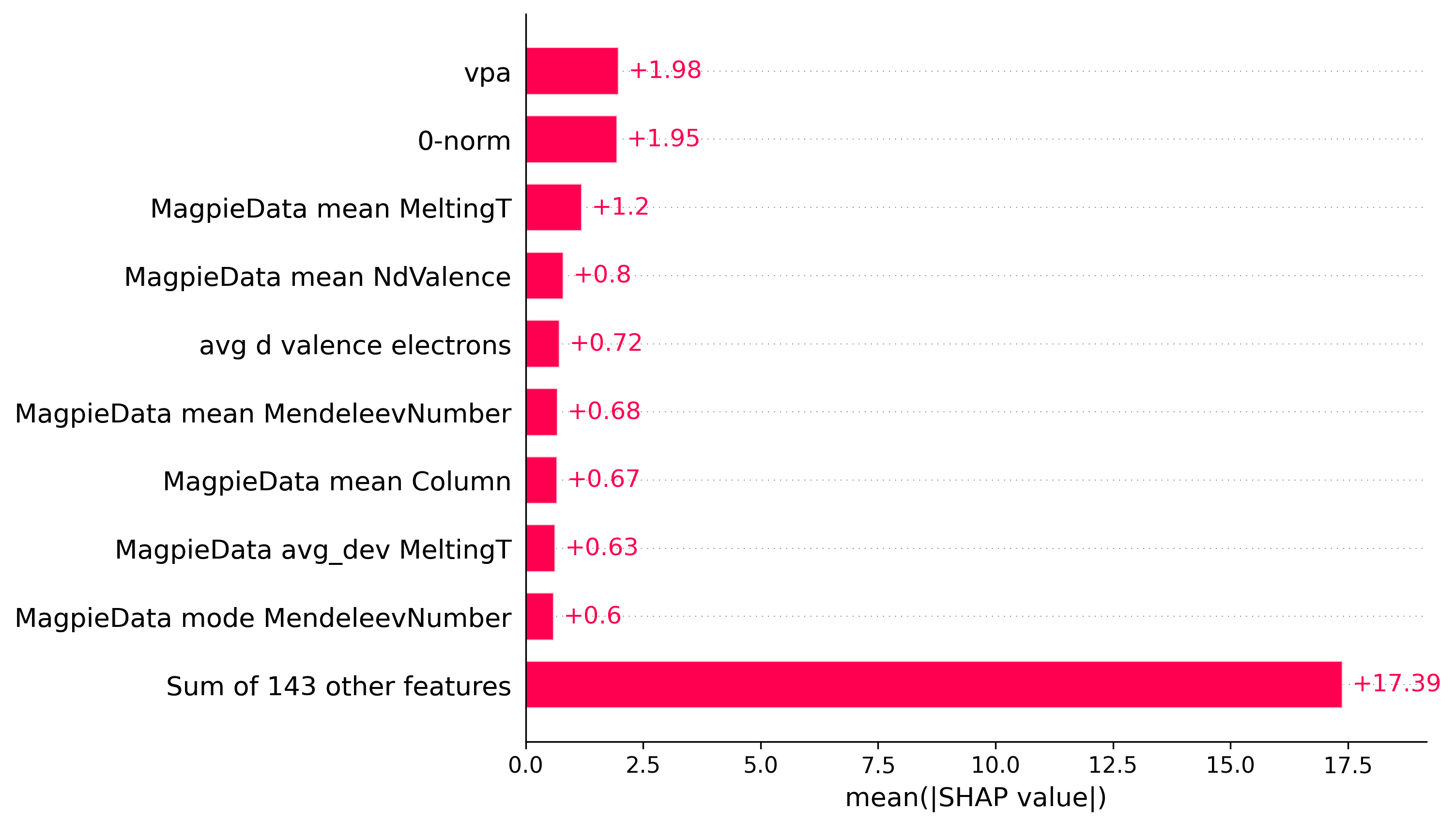}
        \caption*{(a) FM SHAP Analysis}
    \end{minipage}
    \hfill
    \begin{minipage}[b]{0.49\textwidth}
        \centering
        \includegraphics[width=\textwidth]{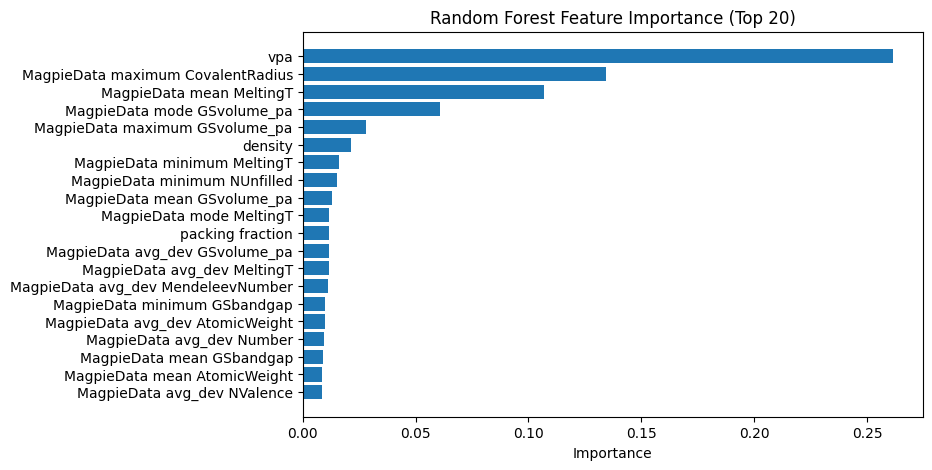}
        \caption*{(b) RF Feature Importance (Impurity)}
    \end{minipage}
    \caption{
    Comparison of feature attribution between the Foundation Model (FM) and Random Forest (RF) for thermal conductivity prediction.  
    (a) SHAP analysis of the FM highlights distributed attention over many correlated descriptors, including lattice stiffness (MeltingT), atomic volume (vpa), and compositional complexity (0-norm).  
    (b) The RF importance ranking shows a hierarchical focus on a few dominant variables such as NUnfilled and CovalentRadius. Detailed RF SHAP analyses are provided in the Supplementary Figure S3.
    }
    \label{fig:shap_analysis}
\end{figure}

\paragraph*{Model comparison on feature importance to LTC}
The RF model (Figure~\ref{fig:shap_analysis}(b)) exhibits a sharp concentration of feature importance to LTC prediction, 
relying heavily on a small number of independent features such as \texttt{NUnfilled} and \texttt{CovalentRadius}. 
In contrast, the FM (Figure~\ref{fig:shap_analysis}(a))distributes its attributions across a wide set of correlated descriptors, a pattern characteristic of attention-based models that aggregate weak signals into a cohesive physical representation. 
This distributed reasoning allows the FM to encode multiple structural and bonding cues simultaneously.

\paragraph*{Dominant Physical Factors affecting LTC prediction}
Both FM and RF models consistently identify lattice stiffness and atomic packing as the principal determinants of LTC $\kappa$. 
Descriptors such as mean melting temperature (\texttt{MeltingT}) and atomic volume (\texttt{vpa}) capture cohesive strength and packing density, in agreement with the Slack model of phonon transport. 
Higher melting points correspond to stiffer lattices and stronger bonding, leading to larger $\kappa$, whereas increased atomic volume generally softens the lattice and lowers phonon group velocities.

\paragraph*{Compositional Complexity and Electronic Effects related to LTC}
The SHAP analysis of the FM further highlights the \texttt{0-norm} (number of constituent elements) as a key negative contributor for LTC, associating greater chemical disorder with enhanced phonon scattering. 
Meanwhile, the RF’s focus on \texttt{NUnfilled} indicates sensitivity to the metal–insulator transition, separating phonon-dominated from electron-assisted heat transport regimes. 
Overall, the FM yields a more physically consistent and distributed representation of bonding stiffness and disorder effects, while the RF depends on a limited set of hierarchically separated descriptors.

\paragraph*{Hierarchical structural features to LTC prediction via  SHAP Analysis}
We further conducted additional SHAP analyses of the atomic-level, edge-level, and angle-level contributions to the Feature-Fusion FM model (described in Table~\ref{tab:alignn_fm_thermo}). The results are provided in the Supplementary Figure S4).
We find that the atom-level features contribute the most to the final prediction (sum of SHAP values $\approx 10.2$), followed by edge- and angle-level features ($\approx 6.3$ and $5.8$, respectively). 
This hierarchy indicates that the FM integrates information across multiple structural scales, from atomic environments to interatomic bonds and angular correlations to construct a unified and physically consistent representation of thermal transport.

\medskip
While the SHAP analysis can identify which descriptors drive predictions, it does not capture how the ICL-FM reorganizes the latent feature space to reflect thermophysical trends.

\subsubsection*{Interpreting latent representation of ICL-FM via t-SNE Mapping}

\begin{figure*}[htbp]
    \centering
    \subfloat[Sample distribution in terms of raw compositional and structural features.
    Spearmanr $\rho = 0.850$ (Random Forest)]{
        \includegraphics[width=0.48\textwidth]{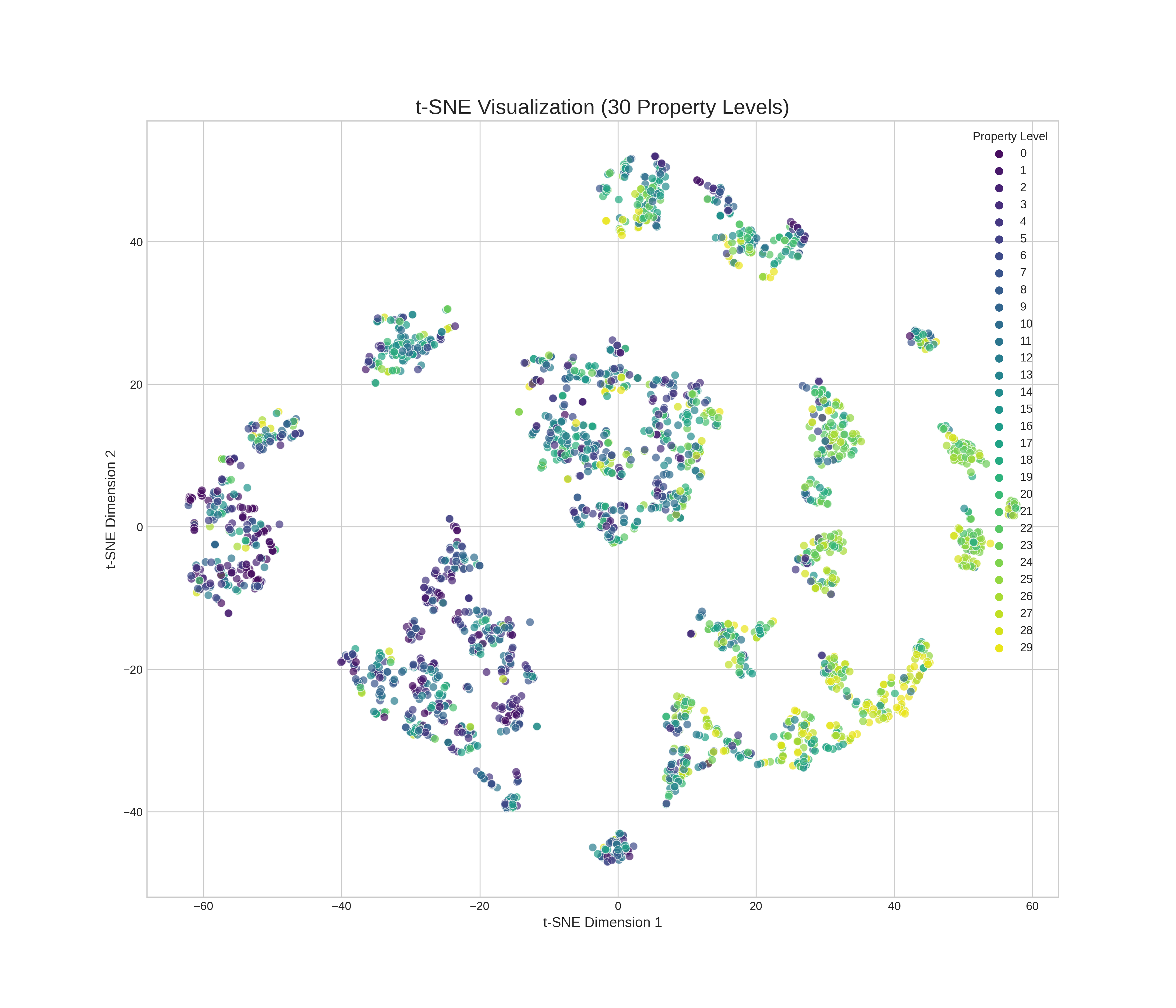}
        \label{fig:tsne_thermo_compo}
    }\hfill
    \subfloat[Sample distribution in terms of FM-learned embeddings from compositional and structural inputs.
    Spearmanr $\rho = 0.925$(ICL-FM)]{
        \includegraphics[width=0.48\textwidth]{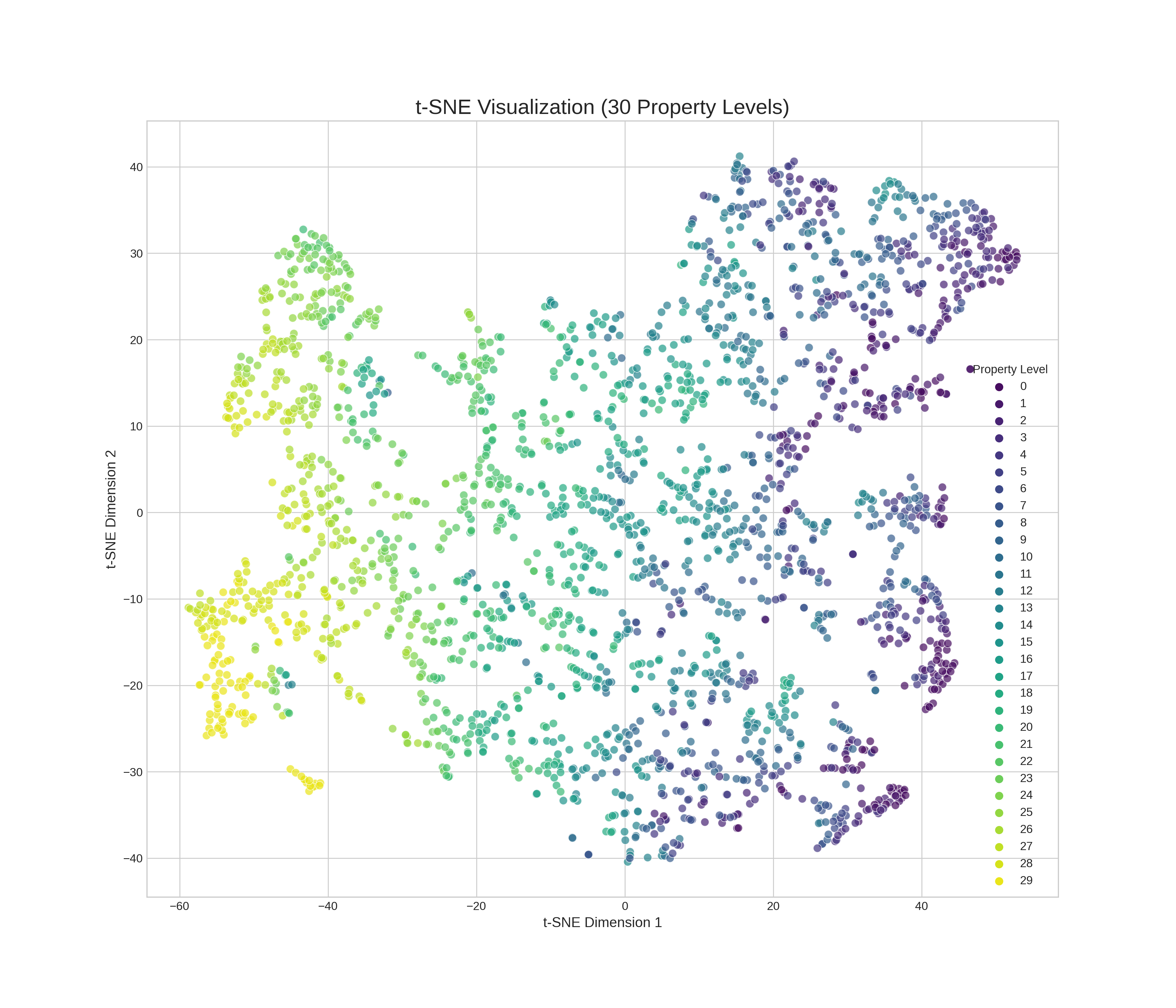}
        \label{fig:tsne_thermo_FM}
    }
    \caption{
    t-SNE visualization of latent representations with LTC  values (fold~4). 
    Each point corresponds to a material, colored by its measured $\kappa$.  
    (a) The raw compositional and structural features form disjoint clusters with weak separation by conductivity.  
    (b) After FM processing, the embeddings become smoother and more continuous, exhibiting clearer gradients with respect to $\kappa$.  
    This transformation indicates that the FM extracts latent physical correlations that align compositional–structural descriptors with the underlying thermal transport physics.  
    }
    \label{fig:tsne_thermo_compo_FM}
\end{figure*}

To visualize how the foundation model (FM) reorganizes compositional and structural features, we employ t-distributed stochastic neighbor embedding (t-SNE) to project the high-dimensional representations generated by the ICL-FM (TabPFN)
onto two-dimensional manifolds colored by lattice thermal conductivity values ($\kappa$). The maps are shown in Figure~\ref{fig:tsne_thermo_compo_FM}.

Figure~\ref{fig:tsne_thermo_compo_FM} (a) shows the material distribution in the t-SNE 2D space projected from the raw composition and structural features. We find that these materials form several disjoint clusters with weak separation by conductivity: high LTC samples (yellow dots) are scattered across several clusters, making it challenging to perform interpolative predictions. This observation is consistent with the relatively lower Spearman rank correlation ($\rho = 0.85$) achieved by the Random Forest model trained on the raw features
In contrast, in the t-SNE map of the ICL-FM embedding space, the materials are distributed as a continuous map, with well-defined trend of low LTC samples on the right while high LTC samples on the left, with clear gradual transitions. 
When predictions are made directly using the ICL-FM, the Spearman rank correlation coefficient increases substantially to $\rho = 0.93$, reflecting a much stronger preservation of the relative ranking of materials by thermal conductivity.
The ICL-FM clearly restructures the raw feature space into a continuous manifold where high and low conductivity materials occupy distinct area of the distribution map. This improvement in rank consistency directly supports the qualitative t-SNE observations and demonstrates that ICL-FM adaptation yields a more well-behaved latent space for accurate ranking and interpolative prediction.
This reorganization reflects the FM’s ability to fuse compositional and structural information into a well-behaviored latent space that leads to high-performance interpolative prediction, capturing relationships among atomic mass contrast, bonding stiffness, and lattice anharmonicity that govern $\kappa$.

\begin{figure*}[htbp]
    \centering

    \subfloat[ALIGNN last-layer features.]{
        \includegraphics[width=0.45\textwidth]{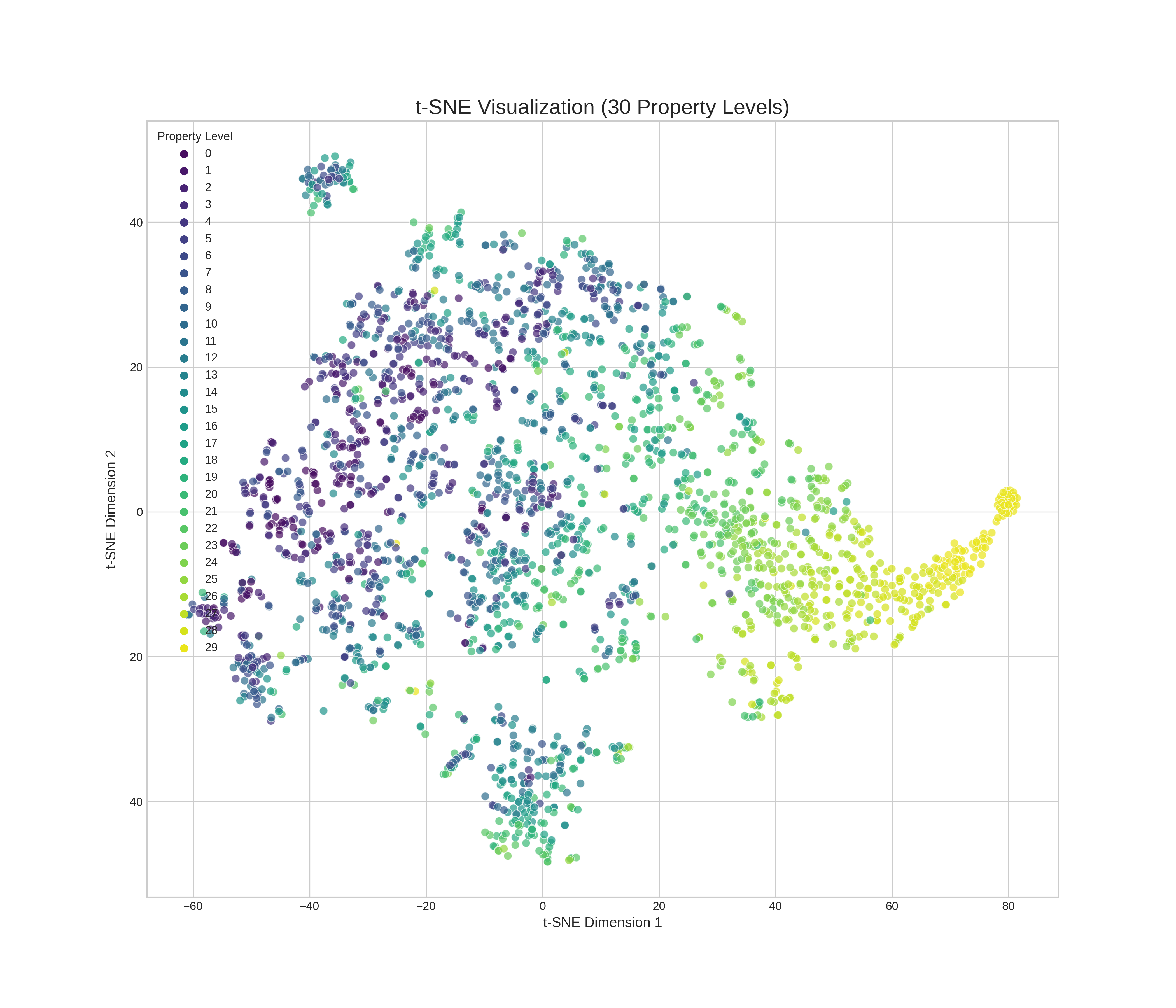}
        \label{fig:tsne_alignn_feat}
    }\hspace{0.03\textwidth}
    \subfloat[FM embeddings from last-layer ALIGNN features.]{
        \includegraphics[width=0.45\textwidth]{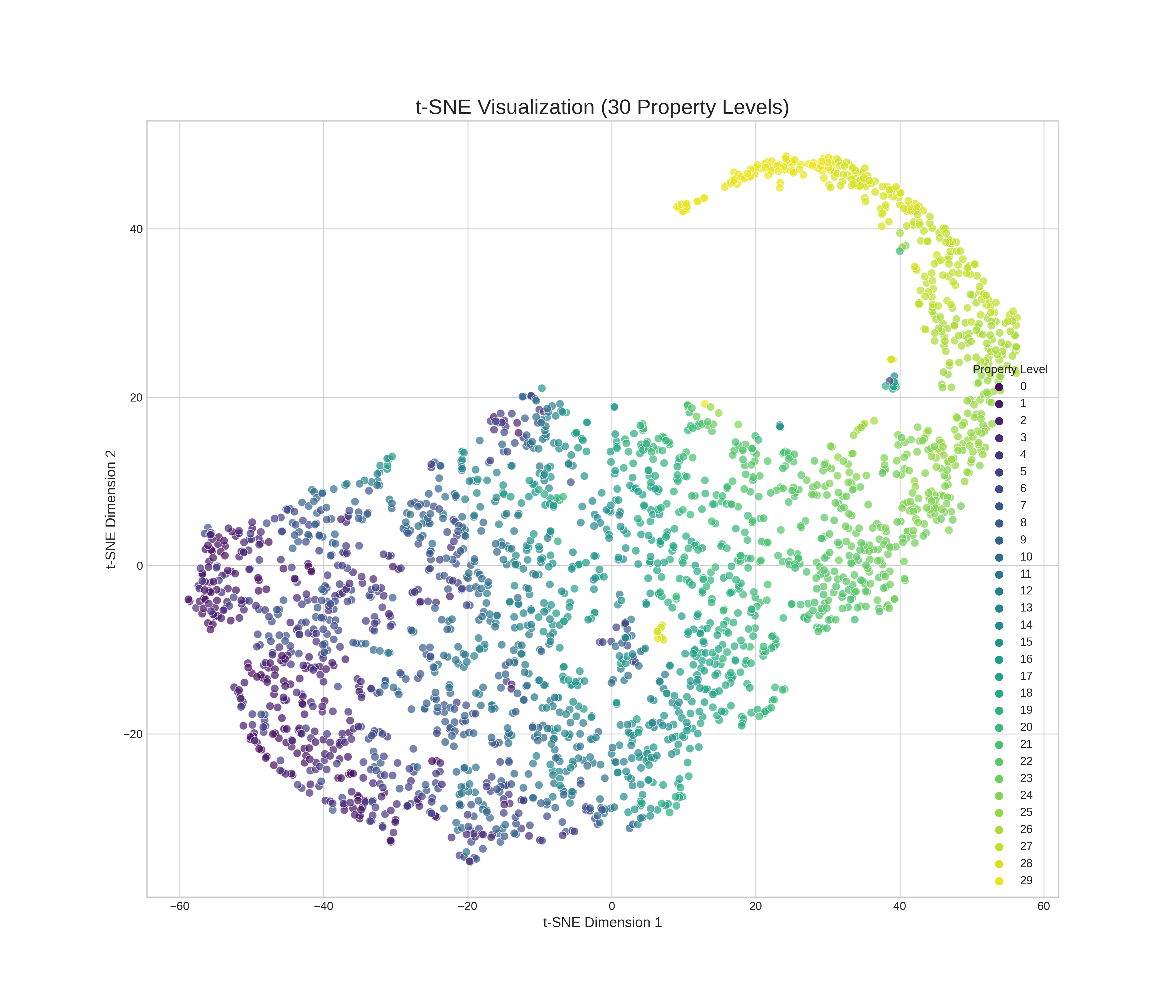}
        \label{fig:tsne_fm_alignn}
    }\\[1.5em]

    \subfloat[Multi-layer fusion features from ALIGNN.]{
        \includegraphics[width=0.45\textwidth]{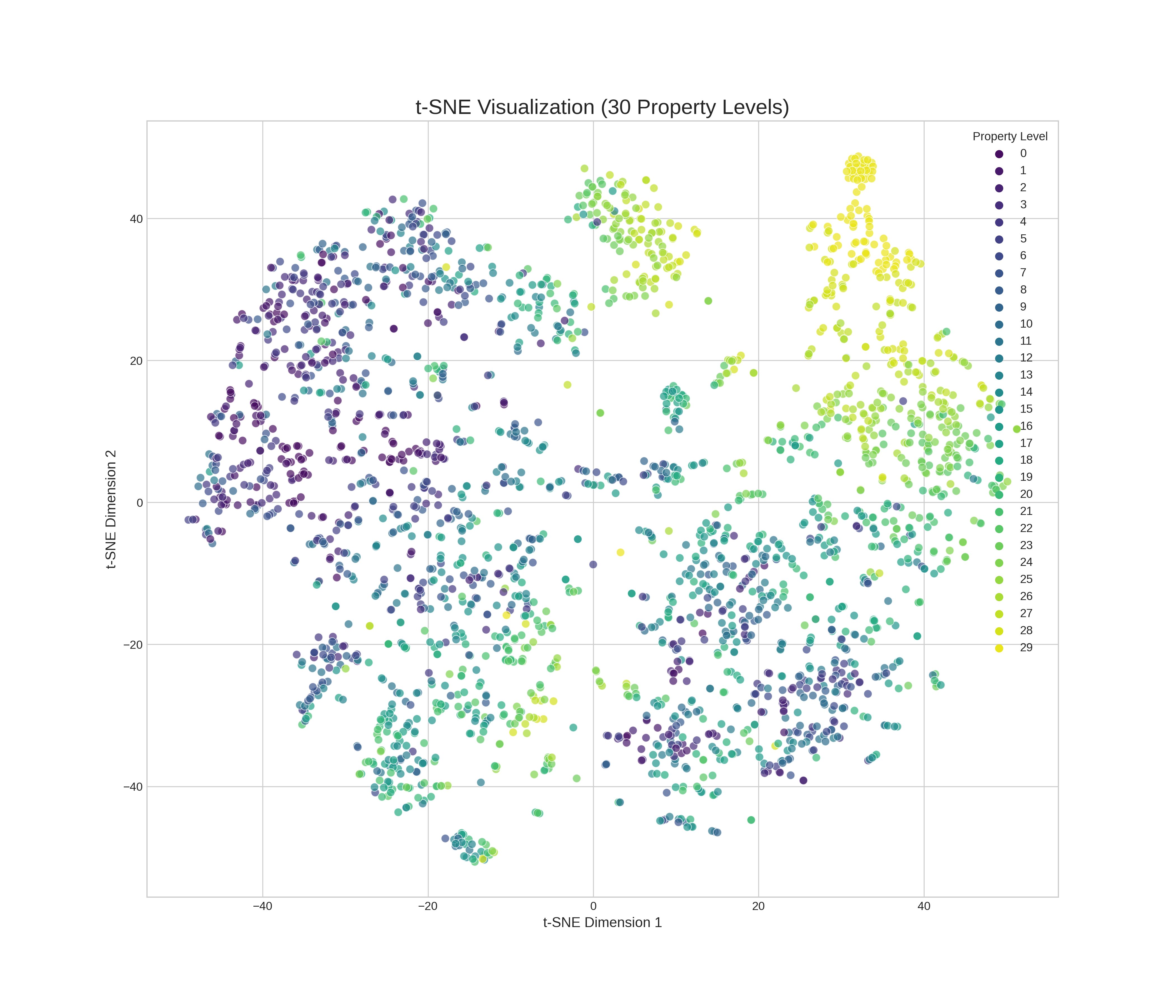}
        \label{fig:tsne_transfer_feat}
    }\hspace{0.03\textwidth}
    \subfloat[FM embeddings from fusion features.]{
        \includegraphics[width=0.45\textwidth]{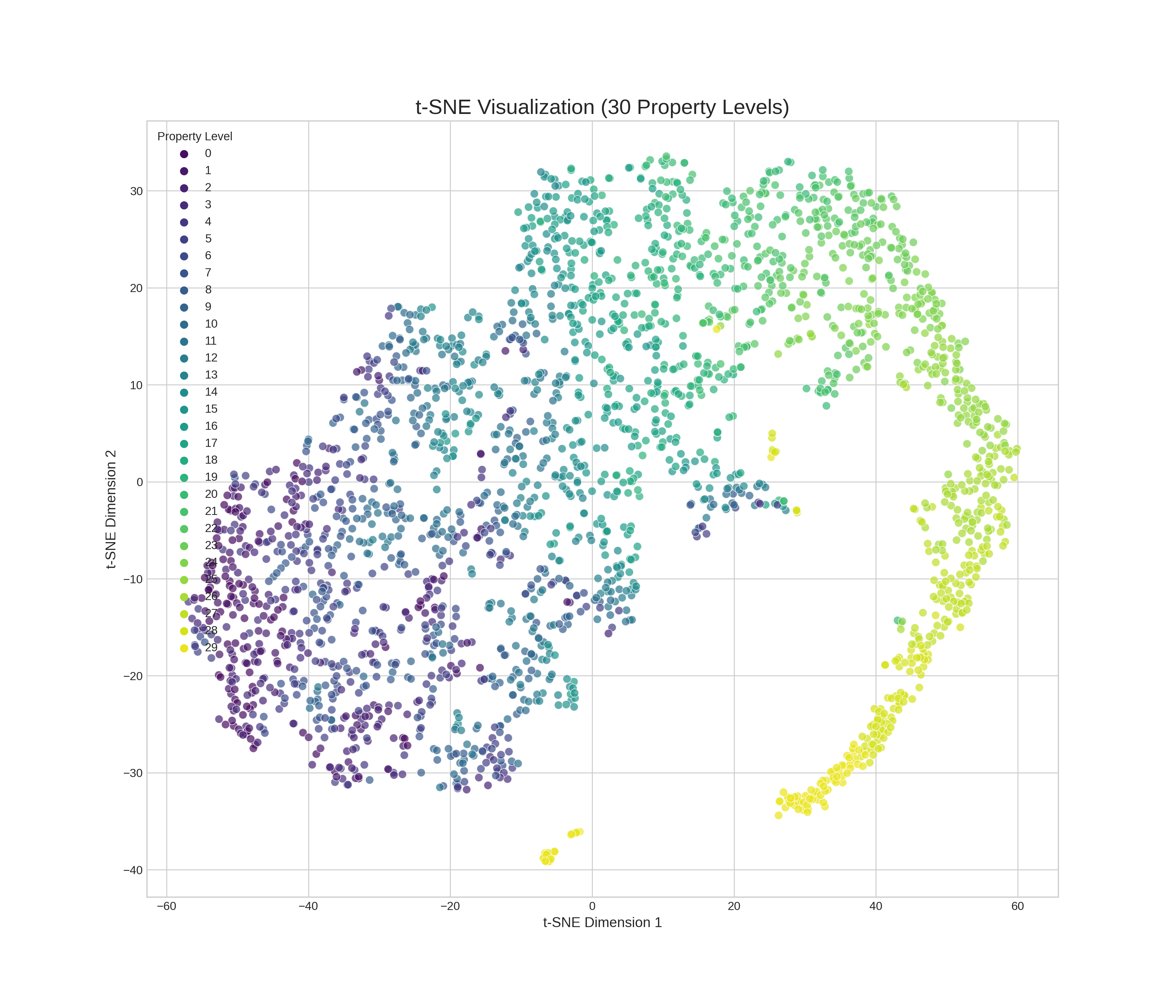}
        \label{fig:tsne_fm_transfer}
    }
    \caption{
    t-SNE visualization of sample distributions in the structural representation spaces for lattice thermal conductivity prediction (fold~4 of our 5-fold cross-validation experiment). 
    (a) Sample distribution in the Last-layer ALIGNN feature space. This feature set capture certain correlations of crystal structures and LTC with overall continuous space. But the continuity of LTC values are not high.
    (b) Sample distribution in the FM-embedding space with Last-layer ALIGNN features: the embedding space become smoother and LTC conductivity is much better than (a). 
    (c) Sample distribution in the Multi-layer fusion feature space, obtained by combining intermediate ALIGNN embeddings. The sample continuity and LTC value continuity are all low.  
    (d) Sample distribution in the FM-embedding space with the fused features. The FM significantly enhances the cluster continuity and alignment with thermal conductivity values.  
    Together, these panels illustrate that the FM acts as a physics-aware feature refiner, improving the structural–property relationship within learned manifolds.
    }
    \label{fig:tsne_alignn_transfer}
\end{figure*}

We also generate t-SNE maps for ICL-FM models with different feature inputs (Figure~\ref{fig:tsne_alignn_transfer}). We find that ALIGNN features already encode structure-dependent correlations as shown by the continuity of sample colors (Figure~\ref{fig:tsne_alignn_transfer}(a)), reflecting its good LTC prediction performance. However, FM models with last-layer ALIGNN features ((Figure~\ref{fig:tsne_alignn_transfer}(b)) and fusion features (Figure~\ref{fig:tsne_alignn_transfer}(d)) produces embedding spaces with smoother and more separable manifolds that better follow the conductivity gradient. 
The sample distribution in the fusion feature space (Figure~\ref{fig:tsne_alignn_transfer}(c))  is much worse compared to the sample distribution in the FM embedding space (Figure~\ref{fig:tsne_alignn_transfer}(d)) after FM transformation of the fusion features. This demonstrates the capability of ICL-FM models to improve the alignment between latent features and physical property (LTC), and amplify the physical coherence of structure–property relationships.

Overall, our t-SNE analyses demonstrate that the FM acts as a physics-aware feature refiner, mapping compositional and structural descriptors onto manifolds that more closely correspond to the LTC landscape. 
The smoother LTC transitions and high continuity of the clusters in the FM-embedding spaces indicate its capacity for in-context learning of complex interactions among bonding topology, mass contrast, and lattice anharmonicity—key factors governing lattice thermal conductivity.

\section*{Discussion}%

This study demonstrates that in-context learning foundation models (ICL-FMs) function as both robust predictive and insightful analytical tools for materials property modeling. By coupling the pretrained TabPFN reasoning engine with representations ranging from handcrafted Magpie features to graph-based ALIGNN embeddings, our framework achieves accurate, data-efficient property prediction without the need for task-specific retraining. Across the MatBench benchmarks and the standalone lattice thermal conductivity (LTC) dataset, the ICL-FM consistently matches or surpasses state-of-the-art (SOTA) models while requiring no fine-tuning effort.

The interpretability analyses emphasize the distinct advantages of this FM-driven approach. Our SHAP and feature-importance studies reveal that the ICL-FM distributes attention across a broad manifold of compositional and bonding descriptors. This contrasts with the hierarchical, node-based feature dependence observed in ensemble methods like Random Forests, allowing the FM to better mirror established physical laws, such as the Slack model of phonon transport. 

The most compelling evidence of the FM's reasoning capability, however, lies in the reorganization of the latent feature space. As revealed by t-SNE analyses, the ICL-FM acts as a physics-aware feature refiner. It transforms "fragmented" feature spaces—where raw descriptors form disjoint clusters with abrupt property changes—into smooth, continuous manifolds where material properties exhibit gradual and physically consistent transitions. 

This shift from discrete clusters to continuous manifolds is physically meaningful, as the smoother latent structure
facilitates interpolation between known materials and improves predictive performance in data-scarce regimes.
At the same time, the gradual transitions observed in the latent space suggest that the model has implicitly captured key physical factors governing material behavior, including the interplay among bonding stiffness, atomic mass contrast, and lattice anharmonicity.
By organizing features along these physically meaningful gradients rather than spurious correlations or noise, the FM exhibits improved generalization and a level of physical coherence that is often lacking in conventional supervised learning approaches trained on limited datasets.

Our thermal conductivity experiments further validate this capacity on an out-of-benchmark dataset requiring coupled chemical and structural reasoning. Despite the absence of domain-specific supervised fine-tuning, the FM demonstrates that it can capture the competing effects of lattice stiffness, compositional disorder, and electronic character that govern phonon-mediated transport. 

Despite its success, we observe that the sample-wise learning capacity of current FMs like TabPFN saturates at approximately 50,000 instances. For massive datasets such as MatBench formation energy or bandgap, task-specific architectures still hold a performance advantage. However, for the small-to-medium datasets that dominate experimental materials science, ICL-FM provides a generalizable paradigm for unifying tabular and graph-based modeling. By treating property prediction as a context-driven inference task, foundation models bridge handcrafted and learned representations to yield interpretable, lightweight, and robust predictors.

\section*{Method}
\label{sec:Method}

\paragraph*{Magpie and Extended Compositional Features}
For composition-based tasks, raw chemical formulas were converted into numerical feature vectors using the standard Magpie featurization pipeline as implemented in the Matminer package. The MagpieEX feature set is implemented in house using Python and Pymatgen Package and is provided with open-source code. The detailed definitions of the four feature sets (Magpie, MagpieEX, Magpie+Struct1, and Magpie+Struct2) 
are provided in the ''Results'' section.

\paragraph*{Baseline Models}
To contextualize the performance of the proposed method, we compare against a diverse set of established baseline models for materials property prediction, spanning both structure-based and composition-based learning paradigms. These include graph neural networks that operate directly on crystal structures as well as attention-based models that rely solely on chemical composition. A brief description of each baseline model and its key architectural characteristics is provided in Supplementary Note S2.

\paragraph*{Structural Features and Graph Embeddings}
In Table~\ref{tab:layer_fm_results}, each embedding corresponds to a specific stage/layer of the ALIGNN network architecture, illustrated schematically in Supplementary Figure S1. 
In Table~\ref{tab:alignn_fm_comparison}, three ALIGNN–FM integration strategies were evaluated: 
\textbf{GNN Input + FM}, \textbf{Last-Layer + FM}, and \textbf{Feature-Fusion + FM}.

\begin{figure}[htbp]
    \centering
    \includegraphics[width=\textwidth]{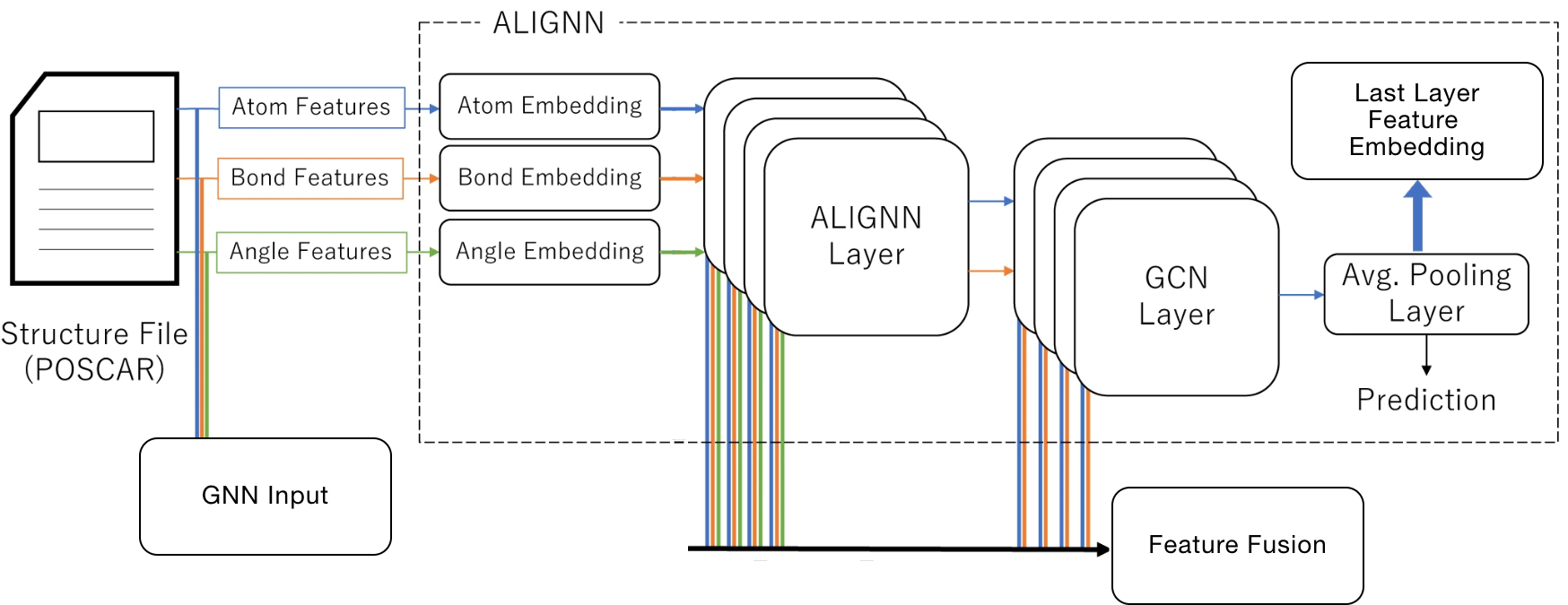}
    \caption{
    Schematic illustration of the three ALIGNN–FM integration strategies used in Table 4 of the main text.  
    (a) \textbf{GNN Input + FM:} raw structural tensors containing atomic indices and neighbor distances.  
    (b) \textbf{Last-Layer } foundation model trained on ALIGNN readout embeddings.  
    (c) \textbf{Feature-Fusion + FM:} multi-layer concatenation of ALIGNN embeddings before FM adaptation.  
    These configurations evaluate how different structural hierarchies contribute to foundation-model learning.
    Adapted and modified from Gupta \emph{et al.}~\cite{Gupta2024Structure}.
    }
    \label{fig:alignn_input}
\end{figure}

Figure~\ref{fig:alignn_input} provides a schematic overview of the representations listed in Table ~\ref{tab:alignn_fm_comparison}.

 The \textbf{GNN Input} representation was generated as follows. Raw crystal structures were converted into fixed-size tabular tensors inspired by the ALIGNN framework. Each structure was encoded as a tensor of shape $[N, 1+2M]$, where $N$ is the maximum number of atoms per unit cell (set to 16) and $M$ is the maximum number of nearest neighbors per atom (set to 12). Each row begins with the atomic number, followed by the indices of its $M$ nearest neighbors (within an 8~\AA{} cutoff) and their corresponding pairwise distances. Structures with fewer than $N$ atoms or atoms with fewer than $M$ neighbors were zero-padded with \texttt{-1}, ensuring a consistent input dimension for all samples. 
This encoding preserves atomic connectivity and distance relationships while producing a uniform feature structure suitable for ICL-FMs.

\subsection*{Training and Evaluation Protocol}
Following the official \texttt{Matbench} benchmarking pipeline, we ensured fair and reproducible evaluation across all tasks. 
The benchmark provides standardized datasets and fixed train/test splits designed for consistent comparison of materials property prediction models. 
We downloaded all tasks using the \texttt{matbench} Python package (\texttt{matbench.client.MatbenchBenchmark}) and strictly adhered to the predefined splits to avoid data leakage and ensure comparability with prior work.  

Each benchmark task was trained independently on its corresponding training set, and model performance was evaluated using the official Matbench scoring functions. 
All regression tasks were assessed using Mean Absolute Errors (MAEs) and, where applicable, Root Mean Squared Error (RMSE).  
This standardized protocol establishes a consistent and transparent basis for comparing our ICL-FM framework 
with other models reported in the Matbench literature.

The same 5-fold cross-validation protocol was applied to the standalone thermal conductivity dataset to ensure consistency with the MatBench evaluation scheme. 
To enable fair comparison among different models, we fixed the random seeds used for data splitting, ensuring identical training and test partitions across all experiments.  
This standardized approach provides a consistent basis for comparing the ICL-FM framework with baseline and state-of-the-art models across both composition- and structure-based tasks.

\paragraph*{SHAP Analysis}
Feature importance was analyzed using the SHapley Additive exPlanations (SHAP) framework to quantify the contribution of individual features and hierarchical groups to model predictions. 
For each trained model, we computed mean absolute SHAP values using the official SHAP Python package, aggregated across all test folds.  %
This analysis was performed for both composition-based and structure-based models, including hierarchical attribution of atomic-, edge-, and angle-level features in the ALIGNN–FM configuration. 
All visualizations were generated using the \texttt{summary\_plot} and \texttt{bar\_plot} functions from the SHAP library.

\paragraph*{t-SNE Representation Mapping}
To visualize the learned feature spaces, we applied t-distributed stochastic neighbor embedding (t-SNE) to raw features and latent embeddings from both the foundation model (FM) and baseline models. 
The t-SNE projections were computed using the \texttt{scikit-learn} implementation with a perplexity of 30, learning rate of 200, and 5,000 iterations. 
Each data point represents a material sample and is color-coded by its measured or predicted property value, providing an interpretable two-dimensional view of how the FM reorganizes compositional and structural representations in relation to the target property.

\subsection*{Implementation Details \& Computational Environment}
All experiments were implemented in Python using model-specific environments for reproducibility. 
\texttt{PyTorch Geometric} was used for GNNs (CGCNN, ALIGNN), 
\texttt{TensorFlow} for coNGN, and custom transformer utilities for TabPFN. 
Each model’s code, preprocessing, and evaluation scripts are organized in separate GitHub folders with accompanying environment files (\texttt{requirements.txt} or \texttt{environment.yml}).  

\paragraph{TabPFN version}
Since TabPFN is an evolving foundation model, all experiments involving TabPFN were conducted using \textbf{TabPFN v2.2.1}, rather than the original 2022 release, to ensure consistency across evaluations.

TabPFN performs inference within seconds due to its single-pass evaluation mechanism, whereas GNN-based models such as ALIGNN typically require several hours of training on larger datasets. 
All experiments were conducted on a workstation equipped with an NVIDIA RTX~4090 GPU (24~GB VRAM), an Intel i7--8700K CPU, and 32~GB of system memory.

\section*{Data Availability}
All benchmark datasets are obtained via matbench API at \url{https://matbench.materialsproject.org/}.

\section*{Contribution}
Conceptualization, J.H.; methodology,J.H., Q.L., R.D.; software, Q.L., R.D.,JF.H.; Investigation, Q.L., J.H., R.D., JF.H., M.H, S.D., S.S.; writing--original draft preparation, J.H., Q.L., R.D.; writing--review and editing, J.H., Q.L., R.D., M.H, N.M., JF.H., S.D., S.S.; visualization, Q.L, R.D.; supervision, J.H., M.H.;  funding acquisition, J.H.

\section*{Competing interest}

All authors declare no financial or non-financial competing interests. 

\section*{Acknowledgment}
The research reported in this work was supported in part by the National Science Foundation under grants 2311202,2110033 and 2320292. The views, perspectives, and content do not necessarily represent the official views of the NSF.

\bibliographystyle{unsrt}  
\bibliography{references}

\end{document}